\documentstyle[aps,multicol,epsf]{revtex}

\begin{document}
\draft
\title{On dispersive energy transport and relaxation in the hopping regime}
\author{O. Bleibaum, H. B\"ottger}
\address{Institut f\"ur Theoretische Physik, Otto-von-Guericke-Universit\"at\\
Magdeburg, Germany}
\author{V. V. Bryksin, A. N. Samukhin}
\address{A. F. Physico-Technical Institute, St. Petersburg, Russia}
\date{\today}
\maketitle

\begin{abstract}
A new method for investigating relaxation phenomena for charge carriers
hopping between localized tail states has been developed. It allows us to
consider both charge and energy {\it dispersive} transport. The method is
based on the idea of quasi-elasticity: the typical energy loss during a hop
is much less than all other characteristic energies. We have investigated
two models with different density of states energy dependencies with our
method. In general, we have found that the motion of a packet in energy
space is affected by two competing tendencies. First, there is a packet
broadening, i.e. the dispersive energy transport. Second, there is a
narrowing of the packet, if the density of states is depleting with
decreasing energy. It is the interplay of these two tendencies that
determines the overall evolution. If the density of states is constant, only
broadening exists. In this case a packet in energy space evolves into
Gaussian one, moving with constant drift velocity and mean square deviation
increasing linearly in time. If the density of states depletes exponentially
with decreasing energy, the motion of the packet tremendously slows down
with time. For large times the mean square deviation of the packet becomes
constant, so that the motion of the packet is ``soliton-like''.
\end{abstract}


\begin{multicols}{2}

\section{Introduction}
\label{intro}
In recent years much attention has been devoted to the study of relaxation
processes of non-equillibrium charge carriers in strongly localized systems,
where transport proceeds via phonon-assisted hopping,
like photoexited charge carriers in band tails (see e.g.
\cite{D3}-\cite{I10}) and Anderson
insulators (see e.g.\cite{I20}-\cite{I22}). In such systems particular small 
relaxation speeds
are observed. Often the smallness of the relaxation speed is attributed to
interaction effects. However, also in strongly localized non-interacting 
electron systems long lasting relaxation processes are known to be the rule,
and not the exception \cite{D3}.

The theoretical investigation of such relaxation processes is notoriously
difficult, since the system is strongly disordered and always in a transient
state. In most problems one is interested in time scales which are large as
compared to the time a single hop needs. For such time scales all quantities
depend strongly on frequency, also for very low frequencies, so that
the consideration of dispersive transport is vital. On the other hand in
most problems of interest both spatial and energetical disorder exists.
Due to the latter fact the transitions are inelastic. The inelastic
character of the transitions, the relaxation, leads to a flow of energy from
the electron system to the phonon system. Due to disorder this transport of
energy is also dispersive. Therefore, dispersive energy transport and
relaxation are intimately connected. The investigation of the relaxation
process requires both the consideration of dispersive particle transport
{\it and} the consideration of dispersive energy transport.

The intricate physical situation manifests also in the equations, which have
to be considered. Since the transport is inelastic already in the simplest
approximation the investigation requires to find solutions to integral 
equations (see e.g.\cite{I23}- \cite{I25}). Due to the fact that the particle 
moves in 
an energy
dependent density of states the kernel of these integral equations does not
depend on the difference between the site energies only. The traditional
method to handle the situation, the percolation theory, is not applicable
here. The effective-medium methods by Movaghar and coworkers give, as
pointed out  by Movaghar and coworkers, wrong results for systems at low
temperature \cite{I12} \cite{I4}. Furthermore, beyond the Markovian 
approximation the 
derivation of these integral equations represents itself an intricate 
problem. To our knowledge, beside the attempts by Movaghar and
coworkers, numerical investigations (see e.g.\onlinecite{D3},
\onlinecite{I3a}), and physical 
intuitive considerations (see e.g. \cite{D3},), mainly Markovian equations 
have been used (see e.g. \onlinecite{I24}). 

It is the aim of the present paper to fill this gap and to provide a
formalism that can be used for studying relaxation phenoma of strongly
localized, charge carriers far from equillibrium in the hopping regime  
taking
into account both dispersive particle transport and dispersive energy
transport. To this end we focus, for the sake of definitness, on 
relaxation of photoxited, non-interacting, non-equillibrium charge carrires 
in band tails of, e.g. amorphous semiconductors, like amorphous Si:H.
On the first glance, this problem seems to be rather special.
If the number of charge carriers exited is small Fermi-correlation is
negligible, so that one has to cope with linear rate equations.
The linearity of the transport equations is, of course, the basic ingredient
in solving the problem. A closer look on the problems of interest reveals, 
however, that most of the problems can be formulated in this 
way. Clearly, how to achieve linearization depends on the experiment chosen,
but, provided the number of charge carrires exited is small,
always the smallness of the particle number can be invoked. In this case,
the number of particles at a site is certainly not small as compared to its
equillibrium value, but small as compared to unity. After linearization the
structure is, in principle, quite similar as for the case considered here. 

Below we present the derivation of a framework for the consideration of
relaxation phenomena due to phonon-assisted hopping at zero temperature.
We focus on the limit of strong localization, where dispersive effects
are expected to be most pronounced. Here the disorder manifests in a strong
dependence of the transport coefficients on frequency for very low
frequencies. Consequently, in this regime the diffusion propagator can not
be calculated from Markovian transport equations, like those used e.g. in
Ref.\onlinecite{I24}. To simplify our integral equations we use the 
concept of
quasi-elasticity, the particle changes its energy only by a small amount 
by one hop.  

The method is applied to the study of dispersive energy transport.
We find that always two
tendencies are present, if only the density of states not decreases with
increasing energy. First, there is a widening of the packet due to the
statistical spreading, the dispersive energy transport. Second, there is
narrowing of the packet due to decrease of the density of states with
decreasing energy. The overall evolution is determined by the interplay of
these two tendencies. We have studied both tendencies in two limiting cases,
for a particle moving in a constant density of states and for a particle
moving in an exponential density of states. In the first case the impact of
the density of states decrease is absent, so that only the statistical
spreading is present. Here the packet evolves into a Gaussian packet moving
with constant drift velocity in energy space. The packet width increases
with time as $\sqrt{t}$. The other result is obtained for exponential
density of states. Here both tendencies are present. We find, that the
velocity of the packet is strongly slowing down with time. In that case the
mean square deviation of energy eventually becomes time-independent.
Consequently, the motion of the packet in energy space is ``soliton-like''.
The concrete results on the diffusion propagator, its time dependence and its
moments for exponential densities of states are of relevance for
photoluminiscense experiments on amorphous Si:H.
\section{Basic equations}

\label{BE}

We consider photoexited, localized charge carriers in band tails at zero
temperture. After exitation the charge carriers lower their energy by
phonon-assisted hops between localized states. Since $T=0$, only hops from
higher to lower energy occur. In this situation the charge carriers are in
strong non-equilibrium. We assume, that the number of excited charge
carriers is small, and their energies are far from the Fermi level, so that
it is very unlikely that an electron jumps to a site already occupied.
Consequently, we can neglect Fermi-correlation. In this case the electron
transport can be described by the rate equation \cite{1,2}.
\begin{equation}
\frac{dp_m}{dt}=\sum_{m^{\prime }}\left( p_{m^{\prime }}W_{m^{\prime
}m}-p_mW_{mm^{\prime }}\right) \,.  \label{B1}
\end{equation}
In calculating the transition probabilities we assume, that the
electron-phonon coupling strength is weak, so that only one-phonon processes
have to be taken into account. Then, at zero temperature, the transition
probabilities are given by
\begin{equation}
W_{m^{\prime }m}=\Theta \left( \omega -\varepsilon _{m^{\prime
}}+\varepsilon _m\right) \Theta \left( \varepsilon _{m^{\prime
}}-\varepsilon _m\right) W\left( \left| {\bf R}_{mm^{\prime }}\right|
\right) ,  \label{B2}
\end{equation}
where
\begin{equation}
W\left( \left| {\bf R}_{mm^{\prime }}\right| \right) =\nu e^{-2\alpha \left|
{\bf R}_{mm^{\prime }}\right| }.  \label{B3}
\end{equation}
Here  $\alpha $ is 
the inverse of localized
state radius, and $\nu $ is the phonon frequency. 
The energy $\omega$ is the upper bound for the energy transfer by one hop.
Note, that in the materials of interest $\omega$ can be much smaller as the
Debye-frequency, since not all phonons can interact with localized
electrons equally well. Short wave-length phonons are ineffective since
the electron-phonon coupling constant tends to zero for momenta $q$ with 
$q/2\alpha\gg1$. Therefore, the effective upper phonon momentum
is of the order $2\alpha$, and not of the order of the inverse lattice 
constant of the host material. Furthermore, in disordered systems the 
high-energetic phonons are localized, and need not contribute to transport 
by necessity. 

The first step
function in front of the transition probabilities restricts the transitions 
to transitions between sites separated by
at most  $\omega $ in energy space. Thus, it decreases the energy
relaxation speed. In impurity conduction, and in nearly all papers on
relaxation of charge carriers in band tails, this step function is usually
replaced by unity. In impurity conduction this is quite reasonable, since it
is assumed that hops are restricted within narrow stripe near the Fermi
level, which is small compared to Debye energy. In the band-tail problem,
however, we can see no reason for neglecting it in advance.

To calculate the transport quantities of interest, we have to calculate the
diffusion propagator. In order to render the analytical calculations
feasible we introduce continuous coordinates. The change of representation
is defined by:
\begin{equation}
n(\rho )=\sum_mp_m\delta \left( \rho -\rho _m\right) \,,  \label{B4}
\end{equation}
where $\rho =({\bf R},\varepsilon )$ and $\rho _m=({\bf R}_m,\varepsilon _m)$%
. In this representation the rate equation (\ref{B1}) takes the form:
\begin{equation}
\frac{dn\left( \rho \right) }{dt}=\int d\rho ^{\prime }n\left( \rho ^{\prime
}\right) V\left( \rho ^{\prime },\rho \right) ,  \label{B5}
\end{equation}
where $V$ is determined by the equations:
\begin{equation}
V\left( \rho ^{\prime },\rho \right) =\int d\rho _1\eta \left( \rho
_1\right) w_{\rho _1}\left( \rho ^{\prime },\rho \right) ,  \label{B6}
\end{equation}
\begin{equation}
w_{\rho _1}\left( \rho ^{\prime },\rho \right) =W\left( \rho ^{\prime },\rho
_1\right) \left[ \delta \left( \rho _1-\rho \right) -\delta \left( \rho
^{\prime }-\rho \right) \right] ,  \label{B7}
\end{equation}
\begin{equation}
\eta \left( \rho \right) =\sum_m\delta \left( \rho -\rho _m\right) .
\label{B8}
\end{equation}
The Laplace-transformed equation is given by:
\begin{equation}
sn\left( \rho \right) -n_o\left( \rho \right) =\int d\rho ^{\prime }n\left(
\rho ^{\prime }\right) V\left( \rho ^{\prime },\rho \right) ,  \label{B9}
\end{equation}
where $n_o(\rho )=n(\rho ,t=0)$ is the initial condition. We will assume
that $p_m(t=0)$ is a function $p_0({\bf R}_m,\varepsilon _m)$, so that $%
n_o(\rho )=p_0(\rho )\eta (\rho )$.

Equation (\ref{B9}) can be solved using the Green's function method. The
solution is given by:
\begin{equation}
n\left( \rho \right) =\int d\rho ^{\prime }n_o\left( \rho ^{\prime }\right)
\Phi \left( \rho ^{\prime },\rho \right) .  \label{B10}
\end{equation}
The Green's function satisfies the equation:
\begin{equation}
s\Phi \left( \rho ^{\prime },\rho \right) -\int d\rho _1V\left( \rho
^{\prime },\rho _1\right) \Phi \left( \rho _1,\rho \right) =\delta \left(
\rho ^{\prime }-\rho \right) .  \label{B11}
\end{equation}
Note, that, due to probability conservation, the Green's function $\Phi $
and the probability $w_{\tilde{\rho}}$ satisfy the relations:
\begin{equation}
\int d\rho \Phi \left( \rho ^{\prime },\rho \right) =\frac 1s\,,  \label{B12}
\end{equation}
\begin{equation}
\int d\rho w_{\tilde{\rho}}\left( \rho ^{\prime },\rho \right) =0\,.
\label{B13}
\end{equation}

\section{Configuration Average}

\label{CA}

In order to calculate the configuration average, we assume, that the sites
are distributed homogeneously in space. The distribution of site energies is
supposed to be given by some distribution function $p(\{\varepsilon _i\})$.
Accordingly, the average of any quantity, depending on the energy and
positions of sites, is given by:
\begin{equation}
\left\langle A\right\rangle =\int \Pi _m\frac{d{\bf R}_m}{{\cal V}}%
d\varepsilon _mp\left( \left\{ \varepsilon _m\right\} \right) A\left(
\left\{ {\bf R}_m,\varepsilon _m\right\} \right) \,,  \label{C1}
\end{equation}
where ${\cal V}$ is the volume of the system. The application of the
averaging procedure to the structural factor $\eta $ serves, in particular,
as a definition for the density of states, i.e.
\begin{equation}
{\cal N}\left( \varepsilon \right) =\left\langle \eta \left( \rho \right)
\right\rangle \,.  \label{C2}
\end{equation}
Products of the structural factor $\eta $ are averaged according to:
\begin{equation}
\left\langle \eta \left( \rho _1\right) \cdots \eta \left( \rho _n\right)
\right\rangle ={\cal N}\left( \rho _1\right) \delta \left( \rho _1-\rho
_2\right) \cdots \delta \left( \rho _{n-1}-\rho _n\right) \,.  \label{C3}
\end{equation}

\begin{figure}
\epsfxsize=80mm
\epsffile{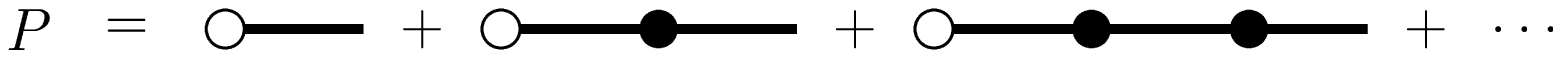}
\caption{ Diagrams contributing to $P$. The leading empty point represents
the single structur factor $\eta $. Everey full dot is associated with a
potential $V(\rho _i,\rho _j)$. Consecutive points are connetcted by $\delta
(\rho _i-\rho _j)/s$, represented by a solid line. Integration is performed
over intermediate arguments.}
\label{FIG1}
\end{figure}

\begin{figure}
\epsfxsize=80mm
\epsffile{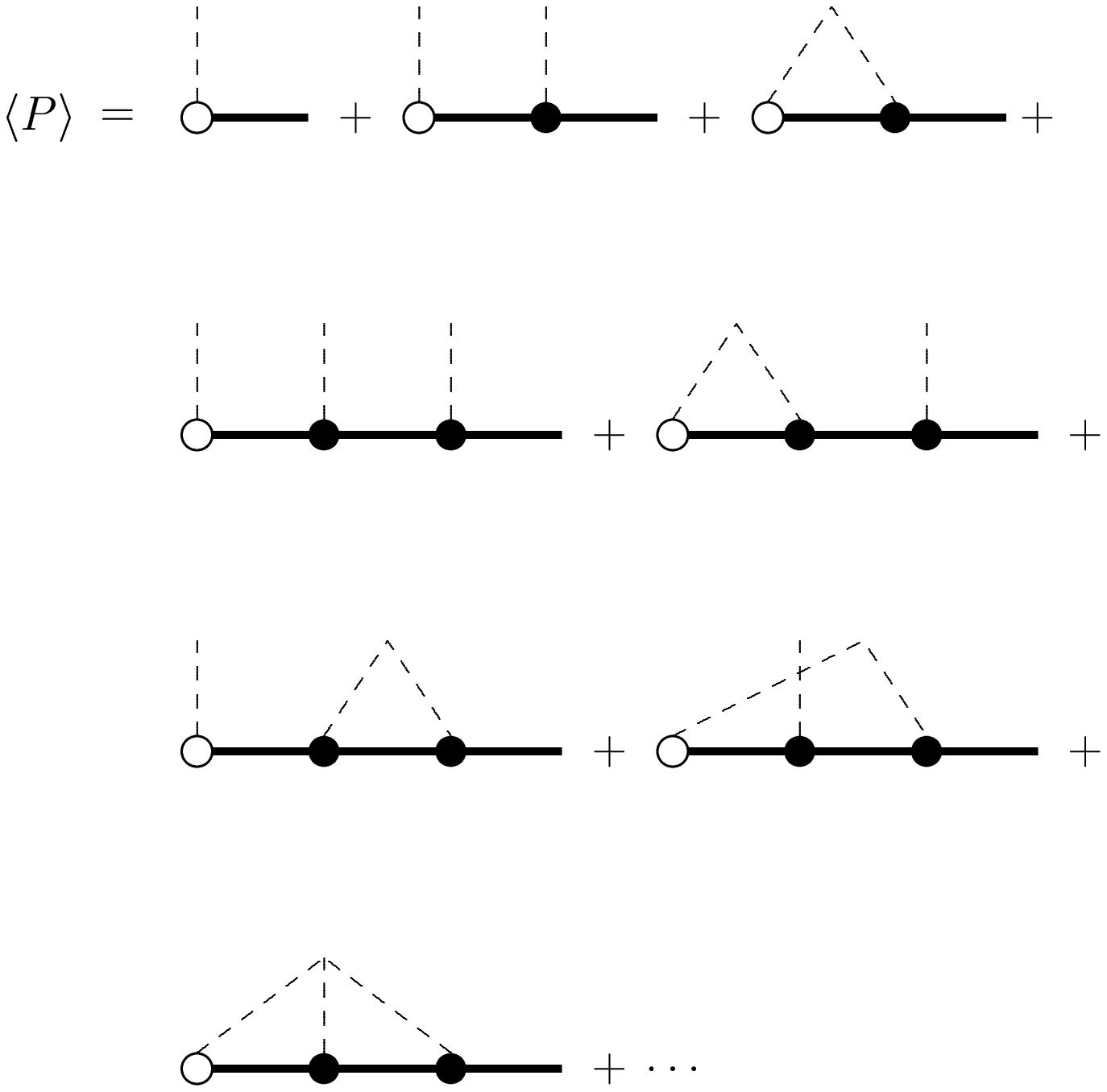}
\caption{Diagrams contributing to the confguration average of $P$. Averaging
is symbolized by dashed lines. Correlated averages are symbolized by joint
lines.}
\label{FIG2}
\end{figure}

\begin{figure}
\epsfxsize=80mm
\epsffile{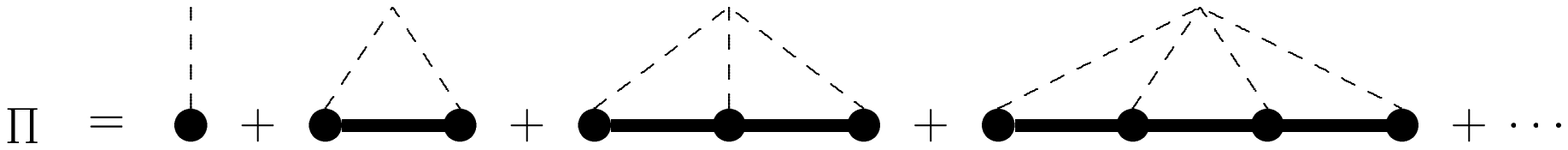}
\caption{Diagrams contributing to $\Pi $. Here, in contrast to Fig.\ref{FIG1}%
, every full line is associated with a function $F(\rho _i,\rho _j)$.}
\label{FIG3}
\end{figure}

\begin{figure}
\epsfxsize=80mm
\epsffile{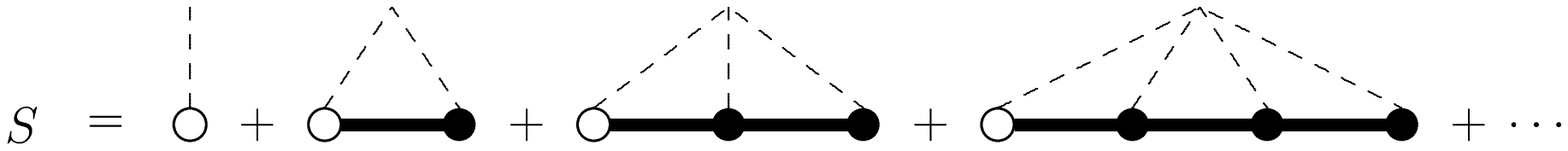}
\caption{Diagrams contributing to S. Here every full line is associated with
a function $F(\rho _i,\rho _j)$.}
\label{FIG4}
\end{figure}

Using these definitions, the configuration average of Eq.(\ref{B10}) can be
calculated diagrammatically \cite{3,4}. The diagrammatic method leads to the
following set of equations for the calculation of the configuration average $%
\langle n(\rho )\rangle $ of the electron density $n(\rho )$ \cite{4},
(see Figs \ref{FIG1}--\ref{FIG4} for illustration):
\begin{equation}
\left\langle n\left( \rho \right) \right\rangle =\int d\rho _1d\rho
_2p_0\left( \rho _1\right) S\left( \rho _1,\rho _2\right) F\left( \rho
_2,\rho \right) \,,  \label{C4}
\end{equation}
\begin{equation}
sF\left( \rho ^{\prime },\rho \right) =\delta \left( \rho ^{\prime }-\rho
\right) +\int d\rho _1\Pi \left( \rho ^{\prime },\rho _1\right) F\left( \rho
_1,\rho \right) \,,  \label{C5}
\end{equation}
\begin{equation}
\Pi \left( \rho ^{\prime },\rho \right) =\int d\rho _1{\cal N}\left( \rho
_1\right) \Pi _{\rho _1}\left( \rho ^{\prime },\rho \right) \,,  \label{C6}
\end{equation}
\begin{eqnarray}
& &\Pi _{\rho _1}\left( \rho ^{\prime },\rho \right) =w_{\rho _1}\left( \rho
^{\prime },\rho \right) \nonumber \\
& + & \int d\rho _2d\rho _3w_{\rho _1}\left( \rho
^{\prime },\rho _2\right) F\left( \rho _2,\rho _3\right) \Pi _{\rho
_1}\left( \rho _3,\rho \right) \,,  \label{C7}
\end{eqnarray}
\begin{equation}
S\left( \rho ^{\prime },\rho \right) ={\cal N}\left( \varepsilon ^{\prime
}\right) \left[ \delta \left( \rho ^{\prime }-\rho \right) +\int d\rho
_1F\left( \rho ^{\prime },\rho _1\right) \Pi _{\rho ^{\prime }}\left( \rho
_1,\rho \right) \right] \,.  \label{C8}
\end{equation}
Here $F(\rho ^{\prime },\rho )=\langle \Phi (\rho ^{\prime },\rho )\rangle $%
. A detailed derivation of the set of integral equations is given in
Appendix \ref{AP1}.

\section{Effective Medium}

\label{EM}

Given the system of integral equations (\ref{C5})-(\ref{C8}), the main
problem is to find an approximate self-consistent solution to it. The
situation is quite similar to that of the calculation of the equilibrium
conductivity in a disordered system \cite{3,4}. There an approximate
solution of this system could be found by introducing a proper decomposition
of the function $F$, the diffusion propagator, into short- and
long-wavelength limit, according to:
\begin{equation}
F\left( \rho ^{\prime },\rho \right) =f(s)C(\varepsilon )\delta \left( \rho
^{\prime }-\rho \right) +\tilde{F}\left( \rho ^{\prime },\rho \right) \,,
\label{EM1}
\end{equation}
where $f(s)$ was a frequency dependent parameter that could be related to
the critical hopping length $R_c$ via the equation:
\begin{equation}
f\nu =\exp \left( -2\alpha R_c\right) \,,  \label{EM2}
\end{equation}
and $C(\varepsilon )$ was an energy dependent function determined by the
principle of detailed balance. This decompostion originated from the notion,
that the integrals in the integral equations are governed by products of
transition probabilities and diffusion propagators, and, for strongly
localized electrons, the latter quantities are short ranged functions as
compared to the transition probabilities. Using the decomposition (\ref{EM1}%
), the effective medium approximation reduces to the repacement of $F$ by $%
fC(\varepsilon )\delta (\rho ^{\prime }-\rho )$ in the calculation of the
effective transition probability $\Pi $ (Eq.(\ref{C7})) and of the
irreducible block $S$ (Eq.(\ref{C8})).

Here in the band-tail problem we use the same philosophy. We first decompose
the diffusion propagator in two parts, according to:
\begin{equation}
F\left( \rho ^{\prime },\rho \right) =f\left( \varepsilon ,s\right) \delta
\left( \rho ^{\prime }-\rho \right) +\tilde{F}\left( \rho ^{\prime },\rho
\right) \,.  \label{EM3}
\end{equation}
Then, to investigate the consequences of this decomposition, we insert Eq.(%
\ref{EM3}) into Eq.(\ref{C7}). Performing a partial summation we obtain:
\begin{eqnarray}
& & \Pi _{\tilde{\rho}}\left( \rho ^{\prime },\rho \right)
={\tilde{w}}_{\tilde{\rho}}\left( \rho ^{\prime },\rho \right) \nonumber \\
&+&\int d\rho _1d\rho _2{\tilde{w}}_{%
\tilde{\rho}}\left( \rho ^{\prime },\rho _1\right) \tilde{F}\left( \rho
_1,\rho _2\right) \Pi _{\tilde{\rho}}\left( \rho _2,\rho \right) \,,
\label{EM4}
\end{eqnarray}
where the renormalized transition probabilities ${\tilde{w}}_{\tilde{\rho}%
}(\rho ^{\prime },\rho )$ are given by Eq.(\ref{B7}), with $W$ replaced by:
\begin{equation}
\tilde{W}\left( \rho ^{\prime },\rho ;s\right) =\frac{W\left( \rho ^{\prime
},\rho \right) }{1+f\left( \varepsilon ^{\prime },s\right) W\left( \rho
^{\prime },\rho \right) }\,.  \label{EM5}
\end{equation}
Note that the renormalized transition probability depends now on $s$ via the
function $f(\varepsilon ,s)$.

At this point the introduction of the function $f(\varepsilon ,s)$, the
effective medium, still seems to be rather arbitrary. However, if we now
choose the effective medium in such a way, that the integrals over $\tilde{F}
$ vanish, the renormalized transition probabilities yield the exact solution
of the diffusion problem. In that case the function $S$ turns into:
\begin{equation}
S\left( \rho ^{\prime },\rho \right) ={\cal N}\left( \varepsilon ^{\prime
}\right) \delta \left( \rho ^{\prime }-\rho \right) \,,  \label{EM8}
\end{equation}
so that Eq.(\ref{C4}) can be cast into the form:
\begin{equation}
\left\langle n\left( \rho ;s\right) \right\rangle =\int d\rho ^{\prime
}\left\langle n_o\left( \rho ^{\prime }\right) \right\rangle F\left( \rho
^{\prime },\rho \right) \,.  \label{EM9}
\end{equation}
Therefore the function $F(\rho ^{\prime },\rho )$ can be identified with the
diffusion propagator.

Note, that although the general philosophy is quite the same as in
equilibrium, the situation is much more intricate here, --- the principle of
detailed balance is absent, since we are dealing with the situation at zero
temperature, and so it can't determine the energy dependence of the
effective medium. Moreover, we expect the critical hopping length to depend
somehow on the position of the electron in the tail. Thus, in contrast to
equilibrium, $f$ is not only a frequency dependent parameter but also an
energy dependent function, which has to be determined self-consistently.

\section{Diffusion propagator in effective medium approximation}

\label{DP}

For the moment we put aside the question of the determination of the
effective medium to elaborate further on the consequences of the
renormalization of the transition probabilities. To this end we focus on the
diffusion propagator.

The equation for the diffusion propagator is given by Eq.(\ref{C1}). In
effective medium approximation, when calculating the irreducible part $\Pi $%
, only the lowest order contribution to $\Pi $ with respect to $\tilde{F}$
is taken into account, so that $\Pi _{\tilde{\rho}}={\tilde{w}}_{\tilde{\rho}%
}$. In that approximation the equation for the diffusion propagator in
momentum representation reads:
\end{multicols}
\widetext
\noindent\rule{20.5pc}{0.1mm}\rule{0.1mm}{1.5mm}\hfill
\begin{equation}
sF\left( {\bf q}|\varepsilon ^{\prime },\varepsilon \right) =
\delta \left(
\varepsilon ^{\prime }-\varepsilon \right) +
\int d\varepsilon _1\left\{
F\left( {\bf q}|\varepsilon ^{\prime },\varepsilon _1\right) \tilde{W}\left(
{\bf q}|\varepsilon _1,\varepsilon ;s\right) {\cal N}\left( \varepsilon
\right) -F\left( {\bf q}|\varepsilon ^{\prime },\varepsilon \right) \tilde{W}%
\left( {\bf 0}|\varepsilon ,\varepsilon _1;s\right) {\cal N}(\varepsilon
_1)\right\} .  \label{P1}
\end{equation}
Of course, as we don't know how the effective medium looks like so far, and
moreover, as the equation is a complicated integral equation, we can't find
a solution. The fact, that the effective medium is a function of energy,
makes the problem much more complicated. Progress can only by achieved if we
can find arguments to simplify the equation considerably. To this end we
focus on the renormalized transition probability. According to the Eqs. (\ref
{B2}) and (\ref{EM5}), it is given by:
\begin{equation}
\tilde{W}\left( R|\varepsilon ^{\prime },\varepsilon ;s\right) =\Theta
\left( \varepsilon ^{\prime }-\varepsilon \right) \Theta \left( \omega
-\varepsilon ^{\prime }+\varepsilon \right) \frac{W\left( R\right) }{%
1+f(\varepsilon ^{\prime },s)W(R)}  
=\Theta \left( \varepsilon ^{\prime }-\varepsilon \right) \Theta \left(
\omega -\varepsilon ^{\prime }+\varepsilon \right) \tilde{W}\left(
R|\varepsilon ^{\prime };s\right) \,.  \label{P2}
\end{equation}
Owing to the step functions in front of the transition probability, the
energy integrations are restricted to intervals of length $\omega $. Taking
this fact into account the integral equation for the calculation of the
diffusion propagator takes the form:
\begin{equation}
sF\left( {\bf q}|\varepsilon \prime ,\varepsilon \right) =\delta \left(
\varepsilon ^{\prime }-\varepsilon \right) 
+\int\limits_0^\omega d\varepsilon _1\left\{ F\left( {\bf q}|\varepsilon
^{\prime },\varepsilon +\varepsilon _1\right) \tilde{W}\left( {\bf q}%
|\varepsilon _1+\varepsilon ;s\right) {\cal N}\left( \varepsilon \right)
-F\left( {\bf q}|\varepsilon ^{\prime },\varepsilon \right) \tilde{W}\left(
0|\varepsilon ;s\right) {\cal N}\left( -\varepsilon _1+\varepsilon \right)
\right\} \,.  \label{P3}
\end{equation}

Further we assume that the diffusion propagator, the transition probability $%
{\tilde{W}}(R,\varepsilon ;s)$, and the density of states ${\cal N}\left(
\varepsilon \right) $ are slowly varying functions on intervals of length $%
\omega $, so that the integrand can be expanded with respect to $\varepsilon
_1$. Doing so we obtain:
\begin{equation}
sF\left( {\bf q}|\varepsilon \prime ,\varepsilon \right) =\delta \left(
\varepsilon ^{\prime }-\varepsilon \right) 
+\omega {\cal N}\left( \varepsilon \right) \left[ \tilde{W}\left( {\bf q}%
|\varepsilon ;s\right) -\tilde{W}\left( {\bf 0}|\varepsilon ;s\right)
\right] F\left( {\bf q}|\varepsilon ^{\prime },\varepsilon \right) +\frac 12%
\omega ^2\frac \partial {\partial \varepsilon }\left[ F\left( {\bf q}%
|\varepsilon ^{\prime },\varepsilon \right) \tilde{W}\left( {\bf q}%
|\varepsilon ;s\right) {\cal N}\left( \varepsilon \right) \right] \,.
\label{P4}
\end{equation}
\hfill\rule[-1.5mm]{0.1mm}{1.5mm}\rule{20.5pc}{0.1mm}
\begin{multicols}{2}
\narrowtext

We terminate the expansion after the first derivative with respect to
energy. This term describes the biased motion of the particle from sites of
higher to sites of lower energy. For finite temperatures one would have to
replace $F\left( {\bf q}|\varepsilon ^{\prime },\varepsilon \right) $ in the
last term with $F\left( {\bf q}|\varepsilon ^{\prime },\varepsilon \right)
+kT\partial F\left( {\bf q}|\varepsilon ^{\prime },\varepsilon \right)
/\partial \varepsilon $ which would describe the energy diffusion current.
At zero temperature there is no input of energy from the phonon-system into
the particle. Therefore, we expect the additional term to be rather small,
and can be neglected.

To investigate the transport properties, we restrict our attention with the
long-wavelength limit. In that case the elastic contribution yields the
diffusion coefficient for particle diffusion. The term, containing the
derivative with respect to energy is finite for $q\to 0$. If we are
interested only in the long-wavelength limit, we can put here $q=0$, since
the remaining terms are of higher order with respect to '$q$' and '$\omega
\partial /\partial \varepsilon $'. Since a nonzero momentum in this term
couples particle diffusion to energy transport, this approximation
corresponds to a decoupling of these two processes. Then, in the
long-wavelength limit, we obtain:
\begin{eqnarray}
sF\left( {\bf q}|\varepsilon ^{\prime },\varepsilon \right) &=&\delta \left(
\varepsilon ^{\prime }-\varepsilon \right) -D\left( \varepsilon ,s\right)
q^2F\left( {\bf q}|\varepsilon ^{\prime },\varepsilon \right)\nonumber \\
&+&\frac \partial
{\partial \varepsilon }\left[ F\left( {\bf q}|\varepsilon ^{\prime
},\varepsilon \right) v\left( \varepsilon ,s\right) \right] \,,  \label{P5}
\end{eqnarray}
where:
\begin{equation}
D\left( \varepsilon ,s\right) =\left. -\frac 12\omega \nabla _q^2\tilde{W}%
\left( q|\varepsilon ;s\right) \right| _{{\bf q=0}}\,,  \label{P6}
\end{equation}
and:
\begin{equation}
v\left( \varepsilon ,s\right) =\frac 12\omega ^2{\cal N}\left( \varepsilon
\right) \tilde{W}\left( 0|\varepsilon ;s\right) \,.  \label{P7}
\end{equation}
Eq.(\ref{P5}) is easily solved. It's solution is:
\begin{equation}
F\left( {\bf q}|\varepsilon ^{\prime },\varepsilon \right) =\frac{\Theta
\left( \varepsilon ^{\prime }-\varepsilon \right) }{v\left( \varepsilon
;s\right) }\exp \left[ -\int\limits_\varepsilon ^{\varepsilon ^{\prime
}}d\varepsilon _1\frac{s+D\left( \varepsilon _1;s\right) q^2}{v\left(
\varepsilon _1;s\right) }\right] \,.  \label{P8}
\end{equation}
Explicit expressions for the transport coefficients can be found in Appendix
\ref{AP2}.

Now, at this stage, the validity of our quasi-elastic approximation
requires, that the second derivative terms are small as compared to terms
with first derivatives with respect to energy. This requirement imposes the
following restrictions on the transport coefficients $v$ and $D$, the
frequency $s$ and the momentum $q$:
\begin{equation}
\omega \left| \frac 1v\frac{dv}{d\varepsilon }\right| \ll 1\,,  \label{P11}
\end{equation}
\begin{equation}
\omega \left| \frac sv\right| \ll 1\,,  \label{P12}
\end{equation}
\begin{equation}
q^2\omega \frac Dv\ll 1\,.  \label{P14}
\end{equation}
The applicability of the quasi-elastic approximation was also discussed
in Ref.\cite{5}. There it was concluded that this approximation should be
inapplicable. To substantiate this statement
numerical calculations were invoked. However, in interpreting these
data it has to be taken into account, that they have been obtained using
a model, neglecting a weighting of the transition probabilities according
to the number of phonons emitted, so that hops between nearly isoenergetic
sites were treated as likely as hops from the very top to the very bottom
of the tail. Consequently the discussion in Ref.\cite{5} applies only to
system with sufficient strong electron-phonon interaction, but not to
systems with weak electron-phonon interaction. For 
extraordinary deep hops to contribute to the diffusion propagator they
should be characteristic for an ensemble of electrons. This is, however,
not expected and in the experimental data, e.g. on amorphous
Si:H \cite{6}, not observed. For these reasons the discussion in
Ref.\cite{5} does not apply to our model.

\section{Physical quantities and interpretation}

\label{PQ}

Before establishing a self-consistency equation, we first elaborate further
on the physical content of our diffusion equation. To this end let us first
have a closer look on the coefficients $D(\varepsilon ,s)$ and $%
v(\varepsilon ,s)$. Imagine, that we have a particle initially located at ($%
{\bf R}_0,\varepsilon _0$). Then the initial condition is ${\langle n_o}(%
{\bf R},\varepsilon )\rangle =\delta ({\bf R}-{\bf R}_0)\delta (\varepsilon
-\varepsilon _0)$. According to our approximation, the motion of the charge
carrier is composed of two contributions: particle diffusion between
isoenergetic sites, and relaxation in energy space. Characteristics of these
two processes are the mean square displacement and the energy relaxation
speed. It turns out, that both can be calculated from the function:
\begin{eqnarray}
P_L\left( \varepsilon _0,\varepsilon ;s\right) =\frac{\Theta \left(
\varepsilon _0-\varepsilon \right) }s\frac \partial {\partial \varepsilon }%
\exp \left[ -s\int\limits_\varepsilon ^{\varepsilon _0}\frac{d\varepsilon _1%
}{v\left( \varepsilon _1,s\right) }\right] \nonumber \\
=\frac{\Theta \left( \varepsilon
_0-\varepsilon \right) }{v\left( \varepsilon ,s\right) }\exp \left[
-s\int\limits_\varepsilon ^{\varepsilon _0}\frac{d\varepsilon _1}{v\left(
\varepsilon _1,s\right) }\right] \,,  \label{PI0}
\end{eqnarray}
which is just $F(q=0|\varepsilon _0,\varepsilon )$.

We define the mean energy of the particle by:
\begin{equation}
E\left( s\right) =\int d\rho ^{\prime }\varepsilon ^{\prime }\left\langle
n\left( \rho ^{\prime },s\right) \right\rangle \,.  \label{PI9}
\end{equation}
Using our diffusion propagator, this equation can be rewritten in the form:
\begin{equation}
E\left( s\right) =\int d\varepsilon ^{\prime }P_L\left( \varepsilon
_0,\varepsilon ^{\prime };s\right) \varepsilon ^{\prime }\,,  \label{PI10}
\end{equation}
and from the diffusion equation we obtain after integration by parts:
\begin{equation}
sE\left( s\right) -\varepsilon _0=-\int d\varepsilon P_L\left( \varepsilon
_0,\varepsilon ;s\right) v\left( \varepsilon ,s\right) \,.  \label{PI11}
\end{equation}
Therefore, in general, the time dependence of the mean energy is given by
complicated integrals. These integrals simplify considerably in two limiting
cases: in the absence of dispersive energy transport, and for energy
independent $v(\varepsilon ,s)$. In the absence of dispersive energy
transport, i.e. in the Markovian limit in which the transport coefficients
are independent of $s$, the integrals can readly be calculated in time
representation. In the latter situation we simply obtain:
\begin{equation}
P_L\left( \varepsilon _0,\varepsilon ;t\right) =\Theta \left( \varepsilon
_0-\varepsilon \right) \delta \left( \varepsilon _m\left( t\right)
-\varepsilon \right) \,,  \label{PI12}
\end{equation}
where $\varepsilon _m(t)$ is defined by:
\begin{equation}
t=\int\limits_{\varepsilon _m}^{\varepsilon _0}\frac{d\varepsilon _1}{%
v\left( \varepsilon _1\right) }\,.  \label{PI13}
\end{equation}
Therefore we obtain:
\begin{equation}
\frac{dE\left( t\right) }{dt}=\frac{d\varepsilon _m\left( t\right) }{dt}%
=-v\left( \varepsilon _m\left( t\right) \right) \,.  \label{PI14}
\end{equation}
Consequently, $v$ is the velocity of energy relaxation, and $\varepsilon
_m(t)$ is the instantenous position of the particle in energy space. In
general, however, in disordered systems the coefficient $v(\varepsilon ,s)$
depends on $s$, so that {\it energy transport is dispersive}. In that case
the integral in Eq.(\ref{PI11}) can only be calculated easily if $v$ is
independent of energy. Then, owing to probability conservation,
\begin{equation}
\left( \frac{dE}{dt}\right) \left( s\right) =-\frac{v\left( s\right) }s
\label{PI15}
\end{equation}
is obtained.

The mean square displacement, defined by:
\begin{equation}
R^2\left( t\right) =\int d\rho ^{\prime }R^{\prime }{}^2\left\langle {n}%
\left( \rho ^{\prime }\right) \right\rangle \,,  \label{PI1}
\end{equation}
can be obtained from similar arguments. In Fourier representation this
equation can be written in the form:
\begin{equation}
\left( \frac{dR^2}{dt}\right) \left( s\right) =-\left. s\Delta _q\right|
_{q=0}\int d\varepsilon ^{\prime }F\left( q|\varepsilon _0,\varepsilon
^{\prime }\right) \,.  \label{PI3}
\end{equation}
The derivative of the diffusion propagator can be calculated using the
diffusion equation. Doing so, we obtain:
\begin{equation}
\left( \frac{dR^2}{dt}\right) \left( s\right) =2d\int d\varepsilon P_L\left(
\varepsilon _0,\varepsilon ;s\right) D\left( \varepsilon ,s\right) \,.
\label{PI5}
\end{equation}
Again Eq.(\ref{PI5}) simplifies only in two special cases, in the Markovian
limit and for energy independent diffusion coefficients. Whereas in the
first case
\begin{equation}
\frac{dR^2}{dt}\left( t\right) =2dD\left( \varepsilon _m\left( t\right)
\right)  \label{PI7}
\end{equation}
is obtained, we have
\begin{equation}
\left( \frac{dR^2}{dt}\right) \left( s\right) =2d\frac{D\left( s\right) }s
\label{PI6}
\end{equation}
in the latter situation. Below we shall see, that energy independent
transport coefficients are obtained for constant density of states only.

\section{The self-consistency equation}

\label{SC}

So far we have only investigated the consequences of the renormalization. In
order to complete the approximation scheme, we still have to calculate the
effective medium itself. Of course, we can not calculate the effective
medium exactly, since this would amount to find an exact solution to the
diffusion problem. Rather we shall try to calculate $f(\varepsilon ,s)$
self-consistently.

The transport coefficients are properties of the Green's function $F$, the
diffusion propagator. Thus, in order to find an equation for $f(\varepsilon
,s)$ we should relate the transport coefficients to $\Pi $, the irreducible
part of the diffusion propagator. The diffusion coefficient comprises only
elastic contributions, however, in the relaxation problem a proper
description of inelastic processes is vital. Moreover, as shown in the
previuos section, the characteristics for particle and energy transport can
already be calculated from the function $P_L$, the $q\to 0$ limit of the
diffusion propagator. Therefore, in establishing a self-consistency equation
we should focus on those characteristics that are important in that limit,
that is the energy relaxation speed $v$.

The equation for the diffusion propagator for $q=0$ is given by:
\begin{eqnarray}
& & sP_L\left( \varepsilon ^{\prime },\varepsilon ;s\right)=\delta \left(
\varepsilon ^{\prime }-\varepsilon \right) +P_L\left( \varepsilon ^{\prime
},\varepsilon ,s\right) \int d\varepsilon _1\Pi \left( \varepsilon
_1,\varepsilon ;s\right)\nonumber\\
& & +\frac{\partial P_L\left( \varepsilon ^{\prime
},\varepsilon ;s\right) }{\partial \varepsilon }\int d\varepsilon _1\left(
\varepsilon _1-\varepsilon \right) \Pi \left( \varepsilon _1,\varepsilon
;s\right) \,.  \label{SC1}
\end{eqnarray}
If we compare Eq.(\ref{SC1}) with Eq.(\ref{P5}) we deduce that
\begin{equation}
v\left( \varepsilon ,s\right) =\int d\varepsilon _1\left( \varepsilon
_1-\varepsilon \right) \Pi \left( \varepsilon _1,\varepsilon ;s\right) \,.
\label{SC2}
\end{equation}
Now we decompose $v$, as defined in Eq.(\ref{SC2}), into two parts, one part
that contains only the effective medium approximation of $\Pi $:
\begin{equation}
v\left( \varepsilon ,s\right) =\int d\varepsilon _1\left( \varepsilon
_1-\varepsilon \right) \Pi |_{EMA}\left( \varepsilon _1,\varepsilon
;s\right) \,,  \label{SC3}
\end{equation}
and a part $\delta v(\varepsilon ,s)$, that contains the deviations $\tilde{F%
}$. Self-consistency requires:
\begin{equation}
\delta v\left( \varepsilon ,s\right) =0\,.  \label{SC4}
\end{equation}
$v(\varepsilon )$, as defined by Eq.(\ref{SC3}), is in accordance with the
definition (\ref{P7}), taking into account Eq.(\ref{T3}) and the inequality
$\omega f^{\prime }(\varepsilon
)/f(\varepsilon )\ll 1$ (Eq. (\ref{P11})), owing to which
contributions proportional to this parameter are negligible.

We now focus on the self-consistency equation (\ref{SC4}). While Eq.(\ref
{SC3}) contains only the effective medium contribution to the diffusion
propagator, $\delta v$ is a functional of the effective medium $f$ and the
deviation $\tilde{F}$. By construction, it is at least linear in $\tilde{F}$%
. If this equation could be solved exactly, an exact solution to the
diffusion problem could be found within quasi-elastic accuracy. In practise
this is not possible, and, therefore, we depend on further approximations.
To simplify this equation, we take into account only the lowest order
contributions to this equation with respect to $\tilde{F}$, i.e. we
linearize $\delta v$ with respect to $\tilde{F}$ and require that the first
order contribution vanishes. This approach is quite close to the usual
CPA-philosophy, in which vanishing of the t-matrix is required in its lowest
order approximation. Using this procedure we obtain the following
self-consistency equation:
\begin{equation}\label{SC5}
\frac{1}{2}\omega^2{\cal N}(\epsilon)f(\epsilon,s){\tilde
W}(0|\epsilon;s)=a\omega-\frac{sb}{{\tilde W}(0|\epsilon;s){\cal N}(\epsilon)},
\end{equation}
where $a$ and $b$ are simply numbers.
A detailed derivation of this equation is given in Appendix \ref{AP3}.

In deriving the self-consistency equations we have imposed further
restrictions on the effective transition probabilities, which determine the
range of its applicability. These
restrictions can be formulated most conviniently using the dimensionless
critical hopping length $\rho _c(\varepsilon ,s)$, related to $f(\varepsilon
,s)$ by (see Appendix \ref{AP2}):
\begin{equation}
f\left( \varepsilon ,s\right) \nu =\exp \rho _c\left( \varepsilon ,s\right)
\,.  \label{SC5a}
\end{equation}
In terms of $\rho _c$, the inequalities (\ref{P11}) and (\ref{P12}), used in
the derivation, read (prime is derivative with respect to $\varepsilon $):
\begin{equation}
\left| \omega \rho _c^{\prime }\left( \varepsilon ,s\right) \right| \ll 1\,,
\label{SC5b}
\end{equation}
\begin{equation}
\left| \rho _c\left( \varepsilon ,0\right) -\rho _c\left( \varepsilon
,s\right) \right| \ll \rho _c\left( \varepsilon ,0\right) \,.  \label{SC5c}
\end{equation}
In addition, when calculating the integrals,
\begin{equation}
\rho _c\left( \varepsilon ,s\right) \gg 1\,,  \label{SC5e}
\end{equation}
was used.

A closed solution to the self-consistency equation can only be found in the 
limit $s=0$. There we obtain:

\begin{equation}
\rho _c\left( \varepsilon ,s=0\right) =\frac{2\alpha }{\left[ \omega {\cal N}%
\left( \varepsilon \right) \right] ^{1/d}}\left[ \frac{2da}{S\left( d\right)
}\right] ^{1/d},  \label{SC6}
\end{equation}
where $S(d)$ is the solid angle in $d$ dimensions.

For $s$ satisfying (\ref{SC5c}), Eq.(\ref{SC5}) can be cast into the form:
\begin{equation}
\left[ \rho _c\left( \varepsilon ,0\right) -\rho _c\left( \varepsilon
,s\right) \right] \exp \left[ \rho _c\left( \varepsilon ,0\right) -\rho
_c\left( \varepsilon ,s\right) \right] =\frac s{\Omega \left( \varepsilon
\right) }\,,  \label{SC8}
\end{equation}
where:
\begin{equation}
\Omega \left( \varepsilon \right) =\frac{2da}{b\omega \rho _c\left(
\varepsilon ,0\right) }v\left( \varepsilon ,0\right) \,.  \label{SC9}
\end{equation}
According to Eq.(\ref{SC8}), the critical hopping length decreases with
increasing frequency. Note, that the structure of the equation (\ref{SC8})
for the calculation of the dispersion of the critical hopping length,
obtained here, agrees completely with that obtained for the critical hopping
length in calculating the equilibrium conductivity \cite{1,4}.

The dispersion of the transport coeffcicients is determined completely by
the dispersion of the critical hopping length. For small $s$ the frequency
dependence preexponetial factors can be ignored, so that from Eq.(\ref{SC8})
explicit equations for the transport coefficients can be obtained. They are
given by:
\begin{equation}
\frac{D\left( \varepsilon ,s\right) }{D\left( \varepsilon ,0\right) }\ln
\frac{D\left( \varepsilon ,s\right) }{D\left( \varepsilon ,0\right) }=\frac s%
{\Omega \left( \varepsilon \right) }\,,  \label{SC10}
\end{equation}
\begin{equation}
\frac{v\left( \varepsilon ,s\right) }{v\left( \varepsilon ,0\right) }\ln
\frac{v\left( \varepsilon ,s\right) }{v\left( \varepsilon ,0\right) }=\frac s%
{\Omega \left( \varepsilon \right) }\,.  \label{SC11}
\end{equation}
The formal solution of these equations is given by the Lambert's W-function $%
{\cal W}(z)$, defined by the equation $z={\cal W}(z)\exp {\cal W}(z)$. Using
the Lambert's function we can write:
\begin{equation}
D\left( \varepsilon ,s\right) =D\left( \varepsilon ,0\right) \exp {\cal W}%
\left( s/\Omega \left( \varepsilon \right) \right) \,,  \label{SC12}
\end{equation}
\begin{equation}
v\left( \varepsilon ,s\right) =v\left( \varepsilon ,0\right) \exp {\cal W}%
\left( s/\Omega \left( \varepsilon \right) \right) \,.  \label{SC13}
\end{equation}
\section{Constant density of states}

\label{cdos}

A constant density of states, although of not much physical relevance, gives
us the unique opportiunity to study pure dispersive energy transport. Here
both $\Omega (\varepsilon )$ and $v_0=v(\varepsilon ,0)$ are independent of
energy. Consequently, Eq.(\ref{PI0}) can readly be integrated. The
integration yields:
\begin{equation}
P_L\left( \varepsilon _0-\varepsilon ,s\right) =\frac{\Theta \left(
\varepsilon _0-\varepsilon \right) }{v\left( s\right) }\exp \left[ -\frac{%
s\left( \varepsilon _0-\varepsilon \right) }{v\left( s\right) }\right] \,.
\label{CDS1}
\end{equation}
The time-dependence of this function can be obtained by the inverse
Laplace-transformation. Using Eq.(\ref{SC11}) to change the integration
variable from $s$ to $y=v/v_0$, the inverse Laplace-transform of Eq.(\ref
{CDS1}) may be written as:
\end{multicols}
\widetext
\noindent\rule{20.5pc}{0.1mm}\rule{0.1mm}{1.5mm}\hfill
\begin{equation}
P_L\left( \varepsilon _0-\varepsilon ,t\right) =\frac \Omega {v_0}\Theta
\left( \varepsilon _0-\varepsilon \right) \int_C\frac{dy}{2\pi i}\frac{\ln
y+1}y\exp \left[ \Omega ty\ln y-\frac \Omega {v_0}\left( \varepsilon
_0-\varepsilon \right) \ln y\right] \,,  \label{CDS2}
\end{equation}
with the properly chosen integration contour $C$. At large enough $t$, that
is for $\Omega t\gg 1$, this expression, using the saddle-point method,
gives simply the Gaussian packet:
\begin{equation}
P_L\left( \varepsilon _0-\varepsilon ,t\right) \approx \Theta \left(
\varepsilon _0-\varepsilon \right) \frac \Omega {v_0}\frac 1{2\sqrt{\pi
\Omega t}}\exp \left[ -\frac{\left( \varepsilon _0-\varepsilon -v_0t\right)
^2}{4v_0^2t/\Omega }\right] \,.  \label{CDS3}
\end{equation}
\hfill\rule[-1.5mm]{0.1mm}{1.5mm}\rule{20.5pc}{0.1mm}
\begin{multicols}{2}
\narrowtext
Now let us consider the time dependence of the energy relaxation. According
to Eq.(\ref{PI15}) the velocity of energy relaxation is:
\begin{equation}
\frac{dE(t)}{dt}=-\int_C\frac{ds}{2\pi i}\frac{v\left( s\right) }se^{st}\,.
\label{CD1}
\end{equation}
Again, the time dependence of the energy relaxation speed can be calculated
using asymptotics. Details of the calculations are presented in Appendix \ref
{AP5}, where it is shown that:
\begin{equation}
\frac{dE\left( t\right) }{dt}\approx -v_0\left\{
\begin{array}{l}
1+\sqrt{\frac e{2\pi }}\frac 1{\left( \Omega t\right) ^{3/2}}\exp \left(
-\Omega t/e\right) \,,\;{\rm as\;}\Omega t\gg 1\,, \\
\frac{\sqrt{2\pi }}{e\Omega t}\left[ \ln \left( e/(\Omega t)\right) \right]
^{-2}\,,\;{\rm as}\;\Omega t\ll 1\,.
\end{array}
\right.  \label{CDS4}
\end{equation}
Note, that the problem of the energy relaxation in the case of constant
density of states ($v(\varepsilon ,s)$ is independent of $\varepsilon $) is
completely equivalent to one of the nonmarkovian charge transport in strong
electric field $E$, when the diffusion (described by the second coordinate
derivate) is totally neglected. One has only to replace $v\left( s\right)
\rightarrow u\left( s\right) E$, $u\left( s\right) $ being the mobility\cite
{0,1}. In the context of dispersive particle transport, the regimes $\Omega
t\ll 1$ and $\Omega t\gg 1$ are the regimes of anomalous and normal
``diffusion'', respectively. The main difference between both is in that
while the sites are ususally distributed homogeneously in space, the density
of states is usually an increasing function of energy.
\section{Exponential density of states}

\subsection{The saddle-point approximation and its break down for 
large times}
\label{AES}
The calculation of the time dependence of the energy distribution function
for an arbitrary density of states on the basis of the Eqs. 
(\ref{P5})-(\ref{P7}), (\ref{SC10}) and
(\ref{SC11}) turns out to be a quite intricate problem. A tool that
can be utilized in tackeling the problem is the saddle point approximation.
How to apply the saddle point approximation for the calculation of the
quantities of interest for an arbitrary density of states is shown in
Appendix \ref{SPA}. Below we focus on the exponential density of states,
which is relevant, e.g., for amorphous Si:H.

We assume that the density of states is given by:
\begin{equation}
{\cal N}(\varepsilon )={\cal N}_0\exp \left( 3\frac \varepsilon \Delta
\right) \,.  \label{AES0}
\end{equation}
To simplify the notations we use the abbreviations:
\begin{equation}
{\bar{\omega}}=\frac {b \omega }{2da}\,,  \label{AES1}
\end{equation}
\begin{equation}
{\bar{\nu}}=\frac{2da^2}b\nu \,.  \label{AES2}
\end{equation}
Then, the Eqs.(\ref{T3}), (\ref{SC6}) and (\ref{SC9}) can be cast into the
form:
\begin{equation}
v\left( \varepsilon ,0\right) \equiv v\left( \varepsilon \right) ={\bar{%
\omega}\bar{\nu}}\exp \left[ -\rho \left( \varepsilon \right) \right] \,,
\label{AES3}
\end{equation}
\begin{equation}
\Omega \left( \varepsilon \right) =\frac{{\bar{\nu}}}{\rho \left(
\varepsilon \right) }\exp \left[ -\rho \left( \varepsilon \right) \right] \,,
\label{AES4}
\end{equation}
\begin{equation}
\rho \left( \varepsilon \right) =\rho _c\left( \varepsilon ,0\right) =A\exp
\left( -\frac \varepsilon \Delta \right) \,,  \label{AES5}
\end{equation}
where:
\begin{equation}
A=\frac{2\alpha }{\left[ \omega {\cal N}\left( 0 \right) \right]
^{1/d}}\left[ \frac{2da}{S\left( d\right) }\right] ^{1/d}\,.  \label{AES6}
\end{equation}
Using these equations it follows from the formulas derived in Appendix
\ref{SPA} that the time dependence of the mean squared deviation and
the time dependence of the mean energy are given by
\begin{equation}
\sigma ^2\left( \varepsilon _m,\varepsilon _0\right) \equiv \sigma ^2\left(
t,\varepsilon _0\right) \simeq {\bar{\omega}}\Delta \left[ 1-\frac{t_0^2}{%
\left( t+t_0\right) ^2}\right] \,,  \label{AES7}
\end{equation}
\begin{equation}
\varepsilon _m\left( t\right) \simeq -\Delta \ln \left[ \frac 1A\ln \left(
\frac{\bar{\omega}}\Delta {\bar{\nu}}\left( t+t_0\right) \right) \right] \,,
\label{AES8}
\end{equation}
where
\begin{equation}
t_0\left( \varepsilon _0\right) =\frac \Delta {{\bar{\nu}}\rho \left(
\varepsilon _0\right) v\left( \varepsilon _0\right) }\,.  \label{AES9}
\end{equation}

Note, that, according to the saddle-point approximation, the distribution is
Gaussian. Furthermore, the dispersion is constant for $t\gg t_0$.
Consequently, for time scales in line with the applicability of the
saddle-point approximation, the motion of the energy packet is
``soliton-like'', that is the packet moves without distortion.

The applicability condition for the saddle-point (\ref{SP8}) requires:
\begin{equation}
\frac{\bar{\omega}}\Delta \rho ^2(\varepsilon _m)\ll 1\,.  \label{AES10}
\end{equation}
Because of $\rho (\varepsilon _m)$ grows with $t$, the saddle-point method,
and, consequently, all the results of this chapter, becomes invalid, at
least at sufficiently large $t$. The physical meaning of the condition 
(\ref
{AES10}) is the following: at $t\gg t_0$ the width of the electron's
energies distribution is $\delta \varepsilon =\sqrt{{\bar{\omega}}\Delta }$.
Because of $\rho \sim \exp \left( -\varepsilon /\Delta \right) $, the
variation of $\rho $ is: $\delta \rho /\rho =\sqrt{{\bar{\omega}}/\Delta }$,
and, because $v\sim \exp \left( -\rho \right) $, we have: $\delta v/v=\delta
\rho =\sqrt{{\bar{\omega}}\rho ^2/\Delta }$. Thus, the condition (\ref{AES10}%
) is just the one of the variation of electron's ``velocity'' across the
distribution is smaller, then the velocity itself.

\subsection{Form of the distribution for large
times}

\label{DLT}

Even if the initial condition are such, that the condition (\ref{AES10}) is
fulfilled, and a gaussian distribution is formed, at some moment of time $%
\rho _m\equiv \rho (\varepsilon _m)$ becomes of the order of $\sqrt{\Delta /{%
\bar{\omega}}}$, which definitely should results in some deviation of the
distribution function from its symmetric, gaussian form. For the exponential
density of states, distribution function in the Laplace representation, Eq.(%
\ref{PI0}), using Eqs.(\ref{SC13},\ref{AES3}-\ref{AES5}), may be written as:
\end{multicols}
\widetext
\noindent\rule{20.5pc}{0.1mm}\rule{0.1mm}{1.5mm}\hfill
\begin{equation}
P_L\left( \varepsilon _0,\varepsilon ;s\right) =
-\frac 1s\rho \frac \partial{\partial \rho }
\exp \left[ -\frac 1{\bar\omega} \int\limits_{\rho _0}^\rho dx
\frac{{\cal W}\left( sxe^x\right) }{x^2}\right] \,
=\frac{{\cal W}\left(s\rho e^\rho \right) }
{{\bar\omega} \rho s}
\exp \left[ -\frac 1{\bar\omega}
\int\limits_{\rho _0}^\rho dx\frac{{\cal W}\left( sxe^x\right) }{x^2}\right]
\,,  \label{DLT1}
\end{equation}
where it was set $\bar{\nu}=1$ and $\Delta =1$ by appropriate choice of time
and energy units. Under the same conditions, which were used in the
derivation of the above formula, the following approximation is valid (see
Appendix \ref{AP6}):
\begin{equation}
\int\limits_{\rho _0}^\rho dx\,\frac{{\cal W}\left( sxe^x\right) }{x^2}
\equiv G\left( s,\rho \right) -G\left( s,\rho _0\right) ;\;\;\;
G(s,\rho ) =\frac 1{\rho ^2}{\cal W}\left( s\rho e^\rho \right) \left[ 1+%
\frac 12{\cal W}\left( s\rho e^\rho \right) \right] +O\left( \frac{{\cal W}}{%
\rho ^3},\frac{{\cal W}^3}{\rho ^3}\right) \,.  \label{DLT2}
\end{equation}
When calculating distribution function at sufficiently large times, the
characteristic values of $s$, giving the main contribution into inverse
Laplace integral become so small, that one can suppose $\left| s\right| \rho
_0\exp (\rho _0)\ll 1$, and: $G\left( s,\rho _0\right) \approx s\rho
_0^{-1}\exp \left( \rho _0\right) $. Under this condition the distribution
function approach initial conditions independent shape: $P(\varepsilon
_0,\varepsilon ;t)\rightarrow \phi (\varepsilon ,t+t_0(\varepsilon _0))$, $%
t_0(\varepsilon _0)=\left({\bar\omega} \rho _0\right) ^{-1}\exp \rho _0$,
\begin{equation}
\phi \left( \varepsilon ,t\right) =\int\limits_{-i\infty }^{+i\infty }\frac{%
ds}{2\pi i}\phi _L(\varepsilon ,s)e^{st}=
-\int\limits_{-i\infty }^{+i\infty }%
\frac{ds}{2\pi is}\rho \frac \partial {\partial \rho }
\exp \left\{ st-\frac 1
{{\bar\omega} \rho ^2}{\cal W}\left( s\rho e^\rho \right) 
\left[ 1+\frac 12{\cal W}%
\left( s\rho e^\rho \right) \right] \right\} \,.  \label{DLT3}
\end{equation}

Let us consider, first, momenta of the distribution at the large times. In
Laplace-representation the momenta are defined by the equation:
\begin{equation}
\chi _{nL}\left( s\right) =\int\limits_0^\infty dt\,e^{-st}\left\langle
\varepsilon ^n\right\rangle \left( t\right) =\int\limits_{\rho _0}^\infty
\frac{d\rho }\rho \left( -\ln \rho \right) ^nP_L\left( \varepsilon
_0,\varepsilon ,s\right) \rightarrow e^{st_0}\int \frac{d\rho }\rho \left(
-\ln \rho \right) ^n\phi _L\left( \varepsilon ,s\right) \,.  \label{DLT4}
\end{equation}

\begin{figure}
\epsfxsize=17cm
\epsffile{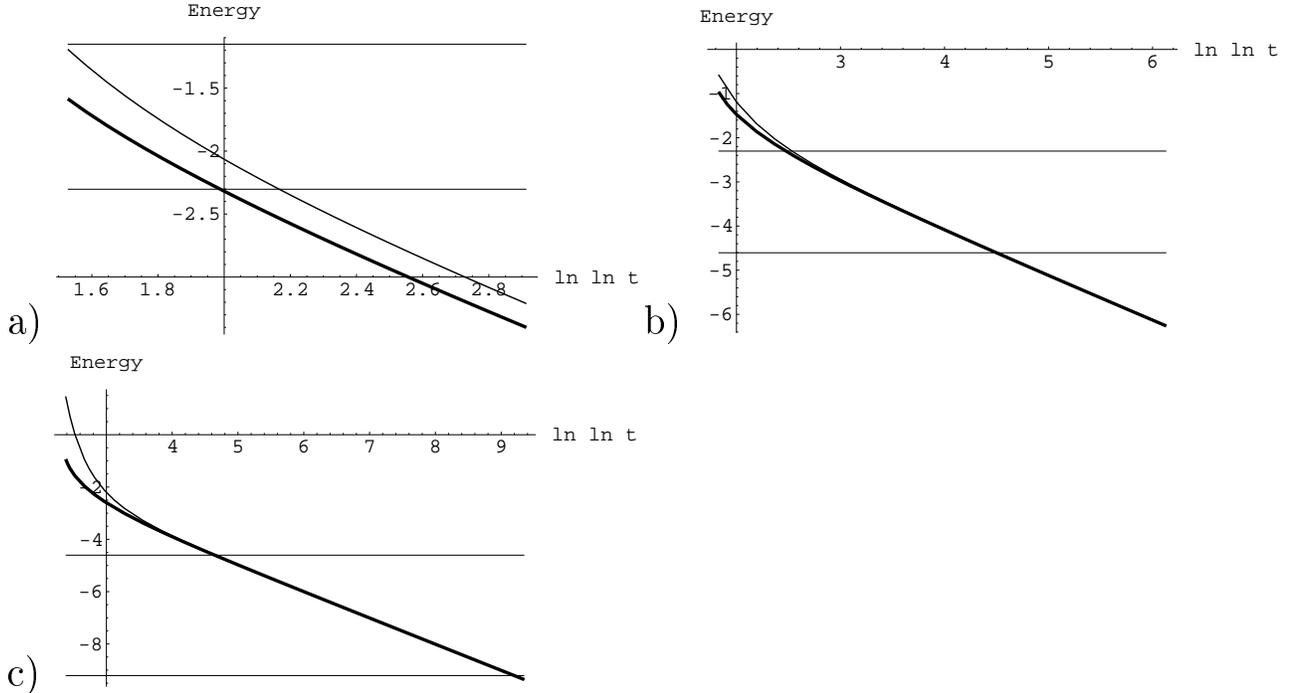}
\caption{Mean energy plotted as a function of $\ln \ln t$ for: (a) ${\bar{%
\omega}}/\Delta =10^{-1}$, $100<t<10^8$; (b) ${\bar{\omega}}/\Delta =10^{-2}$%
, $5\times 10^2<t<10^{200}$; (c) ${\bar{\omega}}/\Delta =10^{-4}$, $5\times
10^4<t<10^{5000}$.}
\label{FIG5}
\end{figure}

\begin{figure}
\epsfxsize=17cm
\epsffile{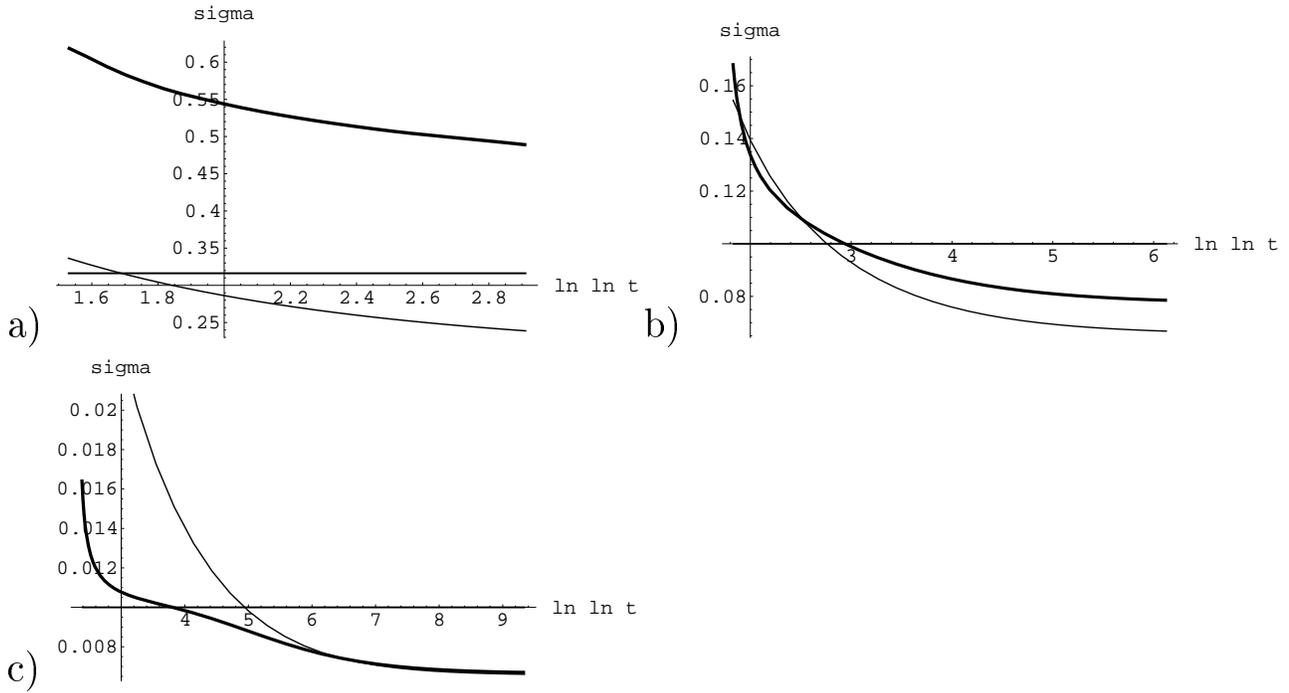}
\caption{Mean square deviation $\sigma $ plotted as a function of $\ln \ln t$
. Values of parameters are the same, as in Fig. \ref{FIG5}.}
\label{FIG6}
\end{figure}

\begin{figure}
\epsfxsize=17cm
\epsffile{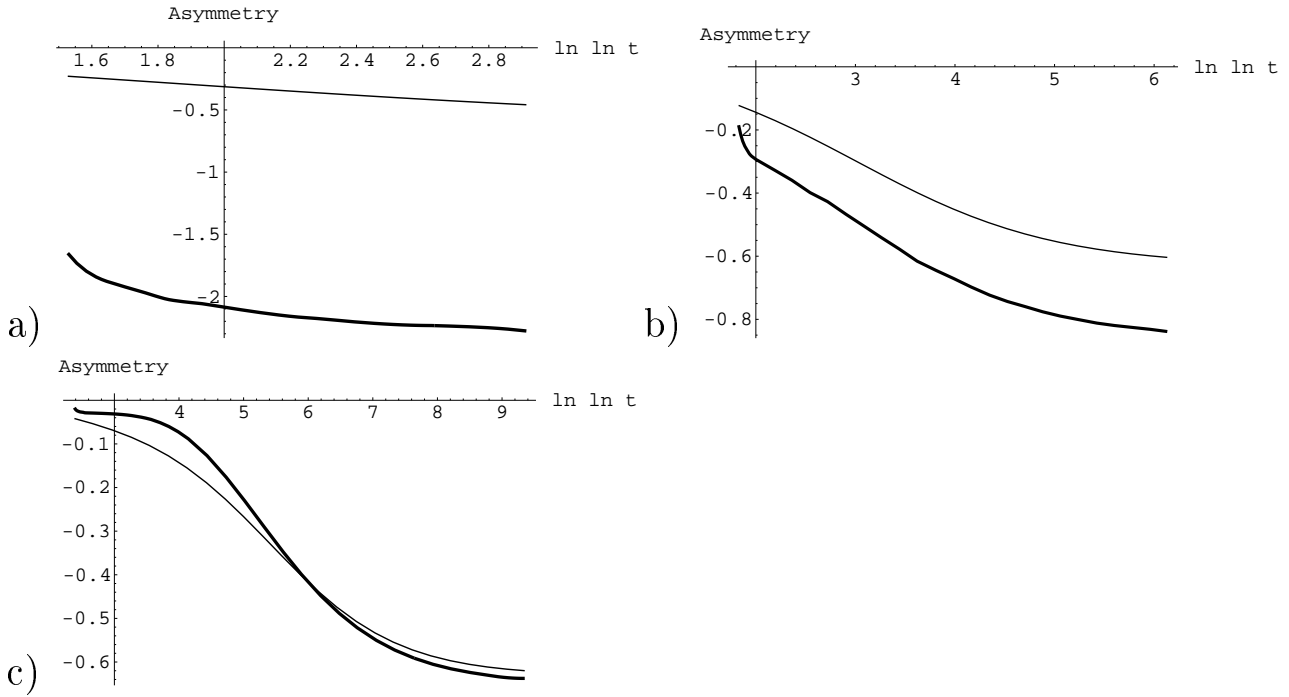}
\caption{Coefficient of asymmetry $\mu _3/\sigma ^3$ plotted as a function
of $\ln \ln t$. Values of parameters are the same, as in Fig. \ref{FIG5}.}
\label{FIG7}
\end{figure}

\hfill\rule[-1.5mm]{0.1mm}{1.5mm}\rule{20.5pc}{0.1mm}
\begin{multicols}{2}
\narrowtext
We shall omit further on irrelevant time shift multiple $e^{st_0}$. It turns
out, that for ${{\bar\omega} }\rho \gg 1$, the integrand of Eq.(\ref{DLT4}) has a
maximum, located at:
\[
\rho _m\approx \left( 1+\sqrt{\bar\omega} \right) \ln 
\frac{\sqrt{\bar\omega}}s%
.
\]
However, if one would try to use the saddle-point method in evaluation of
integrals like (\ref{DLT1}), the results would be incorrect at 
$\sqrt{\bar\omega}\rho _m\gg 1$, since the saddle point criterion is not 
fulfilled.
However, at ${\bar\omega} \rho ^2\gg 1$, momenta can be calculated as an 
expansion
on the powers of small paramenter $1/\left( \sqrt{\bar\omega }\rho \right) $.
Details of the calculations are presented in Appendix \ref{AP6}.

As a result, we have for the first momentum, or the mean energy:
\begin{eqnarray}
&&\chi _1\left( t\right) =\left\langle \varepsilon \right\rangle \left(
t\right) \approx -\Delta \ln \frac 1A\ln \left( \tilde{b}
\sqrt{\bar\omega }%
t\right) \nonumber \\
&&-\sqrt{\frac{\pi {\bar\omega }}2}+\sqrt{\frac \pi {8\bar\omega }}
\frac 1{%
\ln ^2\left( \tilde{b}\sqrt{{\bar\omega }}{t}\right) }+\dots \,,  
\label{DLT5}
\end{eqnarray}
where $\tilde{b}=\sqrt{2}e^{-1+\gamma /2}$, and $\gamma $ is Eulers's
constant. For the distribution's dispersion $\sigma ^2(t)=\mu _2(t)=\chi
_2(t)-\chi _1^2(t)$ we obtain:
\begin{equation}
\sigma ^2\left( t\right) =\mu _2\left( t\right) =\left( 2-\frac \pi 2\right)
{\bar\omega }+\left( 1+\ln 2\right) 
\frac{\sqrt{2\pi \bar\omega }}{\ln \left( \sqrt{%
\bar\omega }t\right) }+\dots \,,  \label{DLT6}
\end{equation}
and for the third central momentum $\mu _3(t)=\left\langle \left(
\varepsilon -\left\langle \varepsilon \right\rangle \left( t\right) \right)
^3\right\rangle =\chi _3(t)-3\chi _1\left( t\right) \chi _2\left( t\right)
+2\chi _1^3\left( t\right) $ we have:
\begin{equation}
\mu _3\left( t\right) =-\left( \pi -3\right) \sqrt{\frac \pi 2}\bar\omega
^{3/2}+3\left[ \pi \left( 1-\ln 2\right) -1\right] \frac{\bar\omega} 
{\ln \left(
\sqrt{\bar\omega }t\right) }\,,  \label{DLT7}
\end{equation}
Figures \ref{FIG5}, \ref{FIG6}, and \ref{FIG7} show mean energy, $\chi _1$,
mean square deviation of energy $\sigma $, and dimensionless coefficient of
asymmetry:
\begin{equation}
{\rm Asymmetry}=\frac{\mu _3\left( t\right) }{\sigma ^3\left( t\right) }\,,
\label{DLT8}
\end{equation}
respectively, plotted versus $\ln \ln t$. There values, obtained
numerically, using the distribution function (\ref{DLT3}), are compared with
their corresponding asymptotic forms, following from Eqs.(\ref{DLT5})-(\ref
{DLT7}).

Moving along the same line, the whole distribution function may be
reconstructed at large times, if to calculate all its momenta. However, it
may be obtained in more simple way. Namely, when $\bar\omega \rho ^2\gg 1$, 
one
may replace $s$ within arguments of ${\cal W}$-functions in Eq.(\ref{DLT3})
with some $s_0=c/t$, $c\sim 1$ (a kind of argumentation may be found in
Appendix \ref{AP7}). Then the integral can easily be calculated to yield:
\begin{eqnarray}
&&\phi (\varepsilon ,t)= \nonumber \\
&-&\rho \frac \partial {\partial \rho }\exp \left\{ -%
\frac 1{\bar\omega \rho ^2}{\cal W}
\left( \frac ct\rho e^\rho \right) \left[ 1+%
\frac 12{\cal W}\left( \frac ct\rho e^\rho \right) \right] \right\} \,.
\label{DLT9}
\end{eqnarray}
\end{multicols}
\widetext
\noindent\rule{20.5pc}{0.1mm}\rule{0.1mm}{1.5mm}\hfill

\begin{figure}
\epsfxsize=17cm
\epsffile{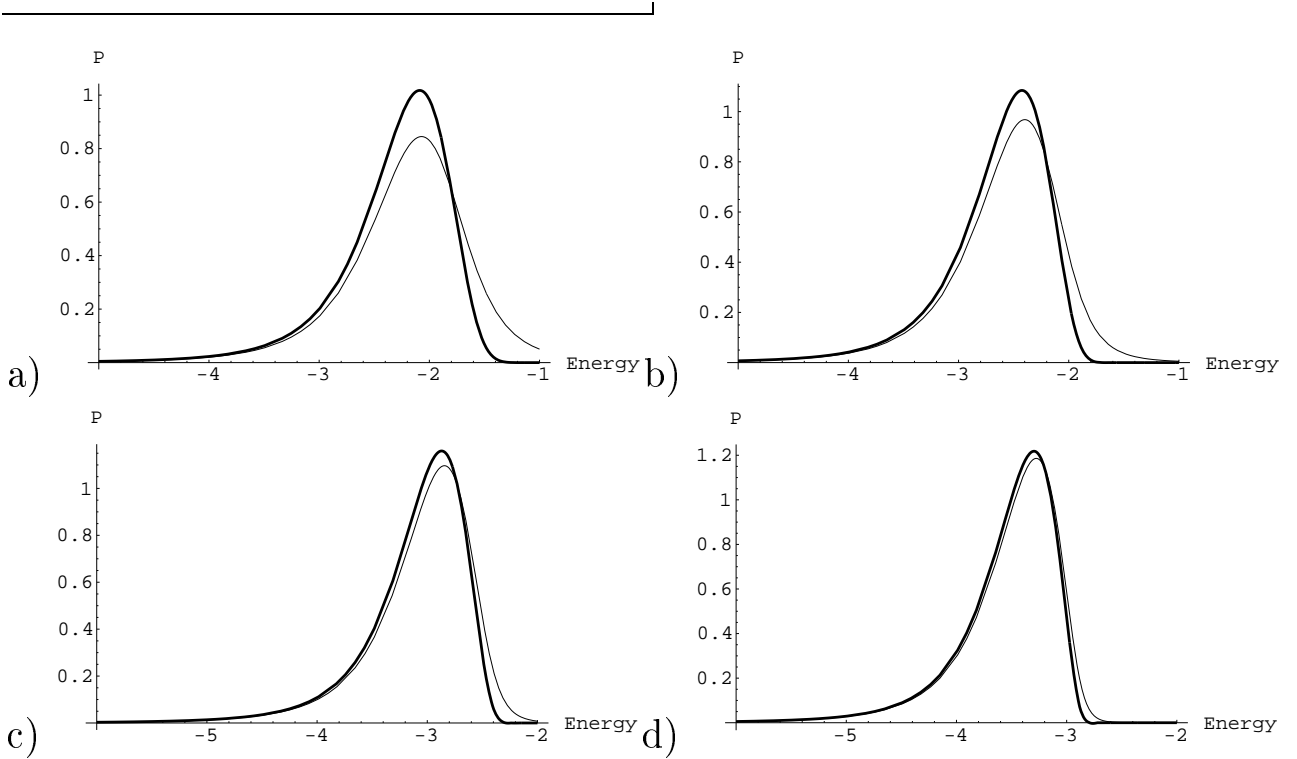}
\caption{Plots of the energy distribution function at ${\bar{\omega}}/\Delta
=0.1$ and: (a) $t=10^3$, (b) $t=10^4$, (c) $t=10^6$, and (d) $t=10^9$.}
\label{FIG8}
\end{figure}

\begin{figure}
\epsfxsize=17cm
\epsffile{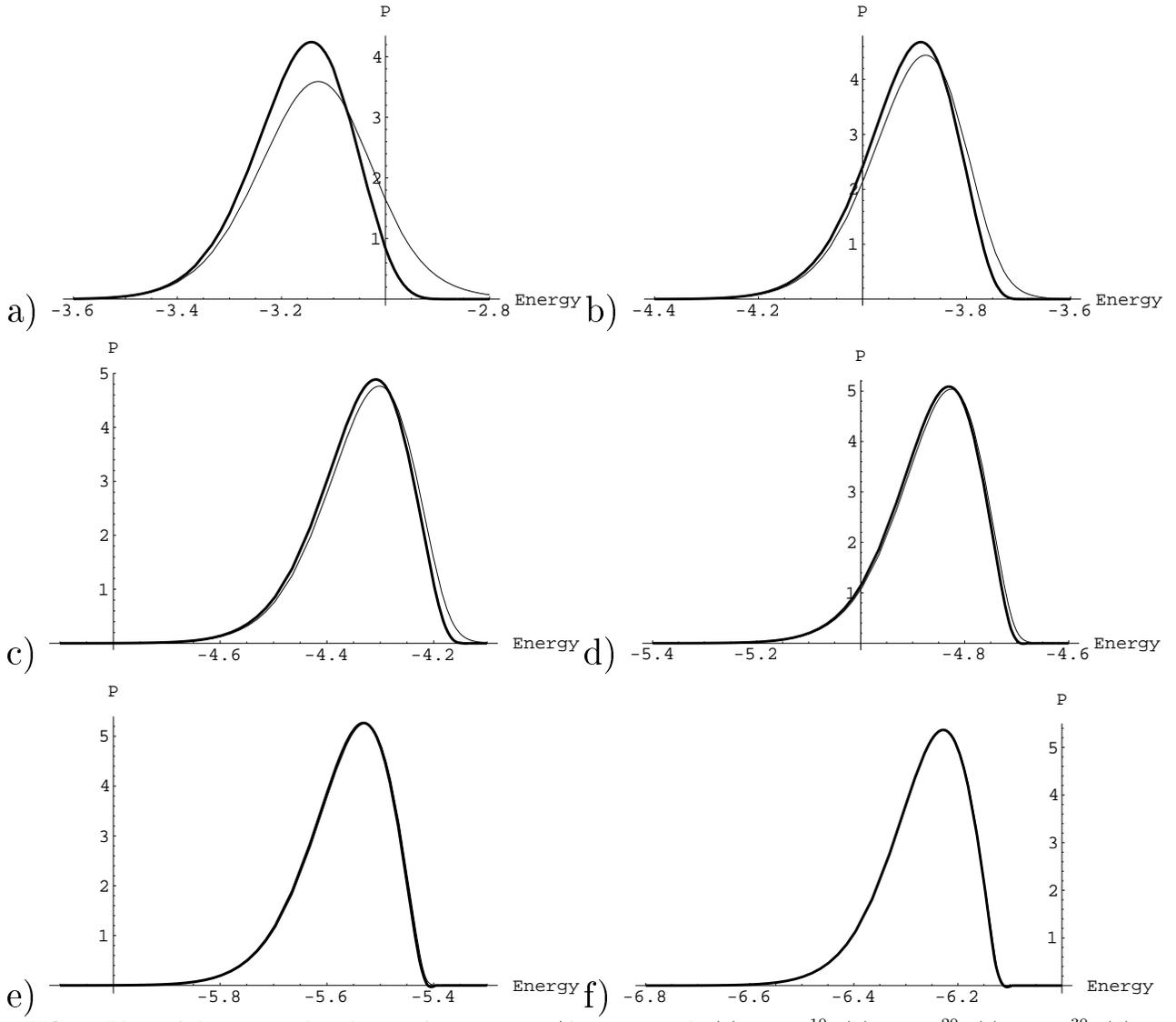}
\caption{Plots of the energy distribution function at
${\bar{\omega}}/\Delta
=0.01$ and: (a) $t=10^{10}$, (b) $t=10^{20}$, (c) $t=10^{30}$, (d) $%
t=10^{50} $, (e) $t=10^{100}$ and (f) $t=10^{200}$.}
\label{FIG9}
\end{figure}

\begin{figure}
\epsfxsize=17cm
\epsffile{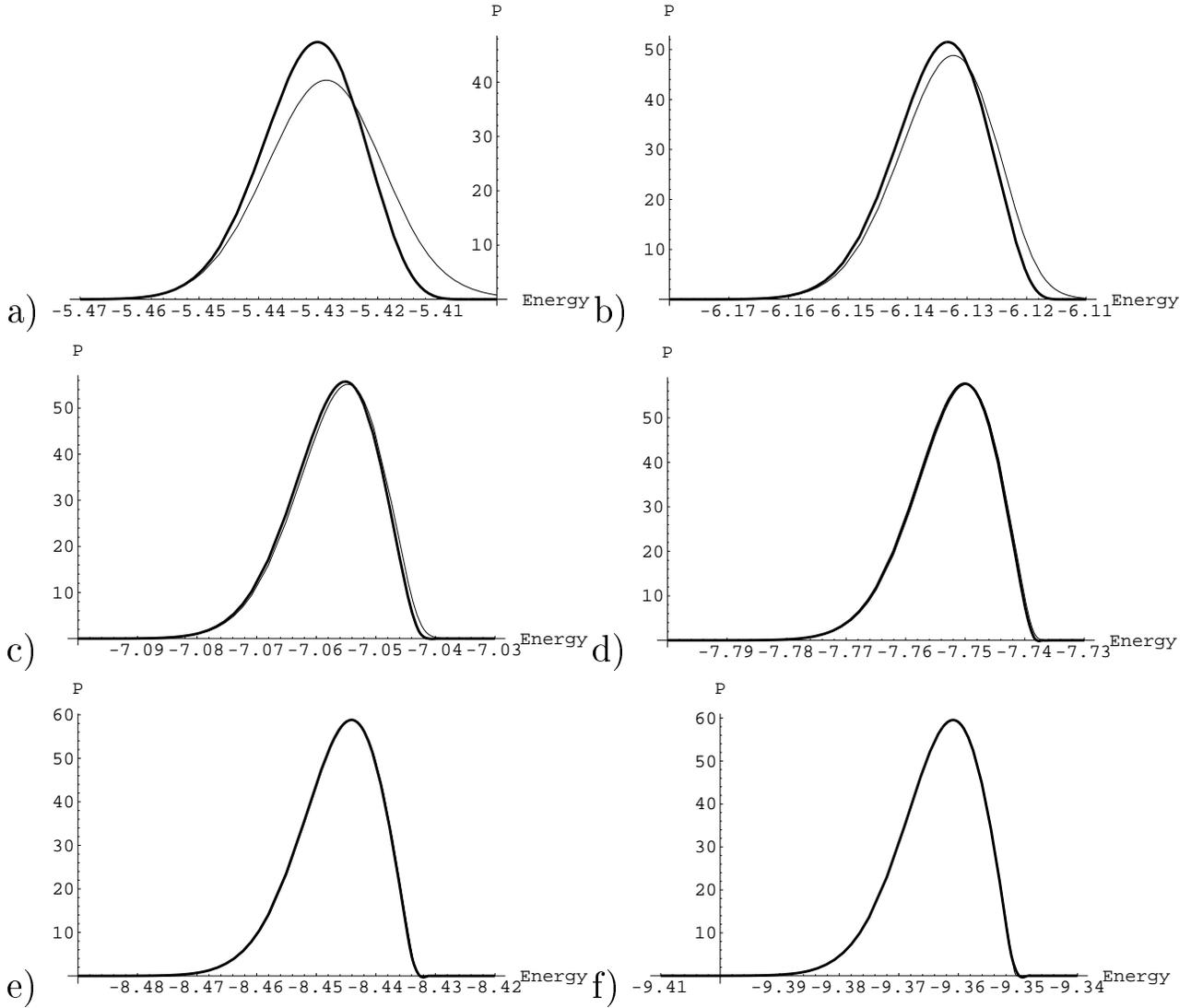}
\caption{Plots of the energy distribution function at $\bar\omega =10^{-4}$ 
and:
(a) $t=10^{100}$, (b) $t=10^{200}$, (c) $t=10^{500}$, (d) $t=10^{1000}$, (e)
$t=10^{2000}$ and (f) $t=10^{5000}$.}
\label{FIG10}
\end{figure}

\hfill\rule[-1.5mm]{0.1mm}{1.5mm}\rule{20.5pc}{0.1mm}
\begin{multicols}{2}
\narrowtext
One can easily show (see Appendix \ref{AP7}), that asymptotic expressions
for momenta (\ref{DLT5}-\ref{DLT7}) may be entirely reproduced from Eq.(\ref
{DLT9}), if to set $c=\exp (-\gamma )$ ($\gamma $ is the Euler's constant).
Figures \ref{FIG8}-\ref{FIG10} show some plots of the distribution function (%
\ref{DLT3}), compared with its asymptotic form (\ref{DLT9}). This latter can
be further symplified, if to introduce $\rho _t$:
\begin{equation}
\frac tc=\rho _te^{\rho _t}\,;\;\rho _t={\cal W}\left( \frac tc\right) \,.
\label{DLT10}
\end{equation}
As we shall see, the body of the distribution at a given time corresponds to
values of $\rho >\rho _t$, $\rho -\rho _t\ll \rho _t$. Then, one can use
approximation: ${\cal W}\left( \left( c/t\right) \rho e^\rho \right) \approx
\rho -\rho _t\gg 1$. Then, we have in the exponent $\left( 1-\rho _t/\rho
\right) ^2/2\bar\omega $, which can be replaced by $\left( \varepsilon
-\varepsilon _t\right) ^2/2\bar\omega $, 
where $\varepsilon _t\approx \ln \ln t$%
. As a result, at large time and small $\bar\omega $ the distribution 
function turns into:
\begin{equation}
\phi \left( \varepsilon ,t\right) =\frac \partial {\partial \varepsilon }%
\left\{
\begin{array}{c}
\exp \left[ -\frac{\left( \varepsilon -\varepsilon _t\right) ^2}
{2\bar\omega }%
\right] \,,\;\varepsilon <\varepsilon _t \\
1\,,\;\varepsilon >\varepsilon _t
\end{array}
\right. \,;\;\varepsilon _t=\ln \ln t\,.  \label{DLT11}
\end{equation}
\subsection{Initial stage of the evolution}

\label{ISE}

At $t=0$ it was supposed $P(\varepsilon ,\varepsilon _0;t=0)=\delta
(\varepsilon -\varepsilon _0)$. To understand, what happens at the intitial
stage of the distribution's evolution, let us consider the behaviour of $P$
at $\varepsilon \approx \varepsilon _0$. One can write from Eqs.(\ref{SP0},%
\ref{SC13},\ref{AES3}-\ref{AES5}):
\begin{eqnarray}
P_0\left( \varepsilon _0,t\right) &\equiv& P\left( \varepsilon
_0,\varepsilon _0;t\right) =\frac 1{\bar{\omega}\rho _0}
\int\limits_{-i%
\infty }^{+\infty }\frac{ds}{2\pi is}{\cal W}\left( \frac s{\bar{\nu}}\rho
_0e^{\rho _0}\right) e^{st} \nonumber \\
&=&\frac 1{\bar{\omega}\rho _0}\int_C\frac{dz}{2\pi
i}\left( 1+z\right) \exp \left( \tau ze^z\right) \nonumber \\
&=&\frac 1{\bar{\omega}\rho _0\tau }\int_C\frac{dz}{2\pi i}\exp \left( \tau
ze^z-z\right)\, ; \nonumber \\
&&\tau =\frac{\bar{\nu}t}{\rho _0}e^{-\rho _0}\,,\;\rho
_0=A\exp \left( -\frac{\varepsilon _0}\Delta \right) \,,  \label{ISE1}
\end{eqnarray}
where the new integration variable 
$z={\cal W}\left( \left( s/\bar{\nu}\right)
\rho _0e^{\rho _0}\right) $ was introduced instead of $s$, and the
integration by parts was performed. This integral may be evaluated using the
saddle-point method. The saddle-point equation, $f^{\prime }(z)=\tau
(z+1)e^z-1=0$, gives us saddle point value $z_c={\cal W}\left( e/\tau
\right) -1$, that is $z_c\approx \ln (1/\tau )-\ln \ln (1/\tau )$ at $\tau
\ll 1$, and $z_c\approx e/\tau -1$ as $\tau \gg 1$. To ensure the
correctness of the saddle point approximation, one should require the
parameter$\left| f^{\prime \prime \prime }(z_c)\right| ^2/\left| f^{\prime
\prime }(z_c)\right| ^3=\left| (3+z_c)^2/(2+z_c)^3\right| $ to be small.
While this is true at $\tau \gg 1$, this parameter appears to be $\approx 1$
at $\tau \ll 1$. Therefore, the result for $\tau \ll 1$ is correct up to the
multiple of the order of 1 only. We have:
\begin{equation}
P_0(\rho ,t)\approx \left\{
\begin{array}{c}
\frac 1{\sqrt{2\pi }\bar{\omega}\rho }\left( \frac e\tau \right) ^{3/2}\exp
\left( -\frac \tau e\right) \text{ as }\tau \gg 1\,, \\
\frac e{\sqrt{2\pi }\bar{\omega}\rho }\left( \ln \frac e\tau -\ln \ln \frac e%
\tau \right) \,\text{as }\tau \ll 1\,.
\end{array}
\right.  \label{ISE2}
\end{equation}

In a similar fasion one can calculate the energy derivative at $\varepsilon
=\varepsilon _0$:
\begin{eqnarray}
&&P_1\left( \rho _0,t\right) \equiv \left. \frac{\partial P}{\partial
\varepsilon }\right| _{\varepsilon =\varepsilon _0}=\left. -\frac \rho \Delta
\frac \partial {\partial \rho }P\left( \rho _0,\rho ;t\right) \right| _{\rho
=\rho _0} \nonumber \\
&=&\frac 1{\bar{\omega}\Delta \rho _0}\int\limits_{-i\infty
}^{+\infty }\frac{ds}{2\pi is}{\cal W}\left( \frac{\Delta {\cal W}}{\bar{%
\omega}\rho _0}+\frac{{\cal W}-\rho _0}{1+{\cal W}}\right) e^{st}  \nonumber
\\
&\approx &\frac 1{\left( \bar{\omega}\rho _0\right) ^2}\int_C\frac{dz}{2\pi i%
}\left[ z\left( 1+z\right) -\frac{\bar{\omega}}\Delta \rho _0^2\right] \exp
\left( \tau ze^z\right) \,,  \label{ISE3}
\end{eqnarray}
where the term $(\bar{\omega}/\Delta )\rho _0z$ within the square brackets
was neglected due to assumption $\left( \bar{\omega}/\Delta \right) \rho \ll
1$. At $\tau \gg 1$ this integral may be readily evaluated in the saddle
point approximation, which yields:
\begin{equation}
P_1\left( \rho ,t\right) \approx -\frac 1{\bar{\omega}\Delta }\sqrt{\frac e{%
2\pi \tau }}\exp \left( -\frac \tau e\right) \,.  \label{ISE4}
\end{equation}
When $\tau \ll 1$, one have to perform integration by parts first, which
gives:
\begin{eqnarray*}
&&P_1\left( \rho ,t\right) =\frac 1{\bar{\omega}^2\rho ^2\tau }\\
&&\int_C\frac{dz}{%
2\pi i}\left[ z-1-\frac{\bar{\omega}\rho ^2}{\Delta \left( 1+z\right) }-%
\frac{\bar{\omega}\rho ^2}{\Delta \left( 1+z\right) ^2}\right] \exp \left(
\tau ze^z-z\right) \,.
\end{eqnarray*}
Then one have in the saddle point approximation:
\begin{eqnarray}
P_1\left( \rho ,t\right) &\approx&
\frac e{\sqrt{2\pi }\bar{\omega}^2\rho ^2}%
\left( z_c^2-1-\bar{\omega}\rho ^2\right) \,, \nonumber \\
z_c &\approx& \ln \frac 1\tau
-\ln \ln \frac 1\tau \,.  \label{ISE5}
\end{eqnarray}
At sufficiently small $t$ the above expression is positive, which means that
the distribution is ``sticked'' near $\varepsilon =\varepsilon _0$, that is,
it monotoneously decreases as $\varepsilon _0-\varepsilon >0$ increases. At
some $\tau =\tau _0$, $P_1$ changes its sign. If $\left( \bar{\omega}/\Delta
\right) \rho _0^2\gg 1$, this corresponds to $z_{c0}\approx \ln \frac 1{\tau
_0}\approx \sqrt{\bar{\omega}/\Delta }\rho _0$, otherwise $\tau _0\sim 1$.
This correspons just to the moment of time, when this distribution separates
from the intial point, --- its maximum is at $\varepsilon <\varepsilon _0$.
For $\tau \ll \tau _0$ the width of the distribution on the energy scale may
be estimated as:
\begin{equation}
\Delta \varepsilon (t)\sim \frac{P_0(\rho ,t)}{P_1(\rho ,t)}\approx \frac{%
\bar{\omega}\rho }{z_c}\approx \frac{\bar{\omega}\rho }{\ln (1/\tau )}\,.
\label{ISE6}
\end{equation}
This formula is valid if $\ln (1/\tau )\ll \rho $, up to the times
corresponding either $\tau \sim 1$ if $\left( \bar{\omega}/\Delta \right)
\rho _0^2\ll 1$, or $\ln \left( 1/\tau \right) \approx \sqrt{\bar{\omega}%
/\Delta }\rho _0$ otherwise. In the former case the distribution width
becomes of the order of $\bar{\omega}\rho $ just before it separates from
the initial point. After this, the distribution, as it was shown in previous
subsection, widens till $\sqrt{\bar{\omega}\Delta }$, and the gaussian
packet moves downward, untill its center reaches the value, corresponding to
$\rho \sim \sqrt{\Delta /\bar{\omega}}$. In the latter case the packet's
width is $\sim \sqrt{\bar{\omega}\Delta }\ll \bar{\omega}\rho $ at the very
moment of its ``unsticking''. In both cases, to consider subsequent
evolution, one has to use some other approximation instead of saddle point
one.
\section{Conclusions}
In the paper we have presented an effective method which permitts the
investigation of relaxation phenomena of localized charge carriers
far from equillibrium due to phonon-assisted hopping at zero temperature.
From the point of view of the formalism the main equations are the
equations for the calculation of the diffusion propagator (\ref{P5}), 
(\ref{P6}) and (\ref{P7}),  and the equations
(\ref{SC9}), (\ref{SC10}) and (\ref{SC11}), which determine the dispersion
of the transport coefficients. These equations show, that both particle
transport and energy transport are dispersive. The equations (\ref{SC10})
and (\ref{SC11}) lead to a strong dependence of the diffusion constant
$D(\epsilon;s)$ and the velocity of energy relaxation $v(\epsilon;s)$
on frequency $s$ already for low frequencies. 
To our knowledge, these equations have not been dervied in the literature
so far, for systems far from equillibrium. In fact, in the literature
mainly frequency independent transport coefficients can be found 
(see e.g. \cite{5} and references therein). The strong dependence 
of the transport coefficients on $s$ results in a non-Markovian equation
for the calculation of the diffusion propagator (\ref{P5}). This
discriminates our equations from the Markovian integral equations used 
in the Refs. \cite{I23}-\cite{I25}. The latter 
equations can, in principle, be obtained
from our leading equation  (\ref{B11}) by neglecting statistical
correlation, so as to average every factor independently.

Using our effective-medium method, we have investigated the relaxation of
charge carriers in band tails at $T=0$. According to our results, energy
relaxation is connected with dispersional transport. Even if we have no real
diffusion in the system, since the temperature is zero, we have some
spreading of the energy distribution with time. For a constant density of
states the situation is completely equivalent to the one in the problem of
the electron's motion in a disordered system, subjected to an electric field
\cite{0}. The main difference between the above mentioned problem and the
energy relaxation is in that, while in the former case the sites are
distributed homogeneously in space, in the energy relaxation problem the
density of states is a decaying function of energy for most physical
systems. In this situation we arrive at the picture that the particles of a
package with lower energies move slower than particles with higher energy.
This leads to the opposite tendency: at first, there is a slowing of the
package spreading. The time dependence of the dispersion, being linear at
the first stage of the evolution, is later slowing down. Later on, different
possibilities exist, dependend on the particular type of the energy
dependence of the density of states.

We performed a detailed investigation for the exponential density of states
for two time regimes. If the variation of the velocity $v$ in the energy
space across the distribution is smaller than the velocity itself, that is
if ${\omega }\rho ^2/\Delta \ll 1$, we arrive at the situation of a package,
moving steadily down along the energy axis without any deformation.
Dispersion becomes time independent (see Eq.(\ref{AES7})). However, since
the particles are sinking down this condition becomes violated when time
goes by. The parameter ${\omega }\rho ^2/\Delta $, being small at the first
stage of the evolution, is getting large. When this parameter is larger than
one, the steady motion condition becomes violated again. The package, being
of gaussian form before, undergoes some restructuring to another,
non-gaussian stable form, with its width lower than before by some numeric
factor of the order of 1 (Eq.(\ref{DLT6})). In our pictures, Figs \ref{FIG7}-%
\ref{FIG10}, it is clearly seen, that the package becomes non-gaussian. This
result remains valid until the very moment the quasi-elasticity conditions
break down ($\omega \rho /\Delta \approx 1$).

For the exponential density of states we have found the packet to move as $%
\ln \ln t$, i.e., its motion is strongly slowing down with time, the packet
becomes, roughly speaking, almost stopped. This type of behaviour may be
called ``glassy'', because of the overall time scale for the packet
evolution (governed by the exponential function of the large parameter $%
\Delta /\omega $), becomes huge. For example, when 
$\omega /\Delta =10^{-4}$, this time scale approaches $10^{1000}$ and 
more (see Fig. \ref{FIG10})!

The main simplification used in our paper is the quasi-elastic
approximation. This approximation relies on the smallness of the
upper bound of the energy transfer to the phonon system by one hop.
For localized electrons, this upper bound can be much smaller as the Debye
energy of the host material, since not all phonons can interact with 
localized electrons equally well. For localized electrons the electron
phonon coupling constant approaches zero, for phonons with wave-vector
$q>2\alpha$. Thus high energetic phonons are less effective. 
Only phonons with energies $\omega<\omega_D2\alpha a$, where $a$ is the
lattice constant of the host material, are effective.
Furthermore,
in disordered systems the high energetic acustical phonons are localized, and
thus need not contribute to transport by necessity. Nevertheless, the
question whether the quasi-elastic approximation is applicable depends much 
on the material of interest. If, however, we compare our result for
the time dependence of the mean energy (\ref{AES8}) in an exponential
density of states with those existing in 
the literature \cite{D3}, we also find agreement for
$\omega/\Delta\approx1$, which indicates that our results are, at least
qualitatively, of wider validity. Unfortunately, to our knowledge, in the
literature there are no further results on the width of the energy 
distribution available to compare with.


From our point of view the main question remained open is how the
results, obtained for the exponential density of states, may be generalized
for other types of energy dependencies. One can imagine, e.g., that for
densities of states decaying with decreasing energy slower than exponential,
the dispersion is growing, but under some sublinear law. Also, in the
opposite case, in a density of states decaying faster than exponential, we
would possibly have a dispersion tending to zero for large times.

It should, however, also be mentioned that it is not completely clear under 
which
situation zero temperature results can be applied to  systems at finite
temperature. In our approach we assumed that $T=0$.
However, in a real system, when the carriers are sinking down, the criterion
for the temperature to be treated as zero, is violated at every finite value
of $T$ at some moment of time, since the contribution of hops to sites with
higher energy values becomes more and more comparable with one of downward
hops. Therefore, the consideration of temperature becomes vital. Work in
this direction is in progress.

\acknowledgments
On of the authors (O.B.) would like to thank Prof. P. Thomas for drawing his
attention to the photoluminscense problem and stimulating discussions.
\end{multicols}
\widetext
\noindent\rule{20.5pc}{0.1mm}\rule{0.1mm}{1.5mm}\hfill

\appendix

\section{On the configuration average}

\label{AP1}

Here we derive the set of integral equations (\ref{C4})-(\ref{C8}). The same
method has already been used in Ref.\cite{4}.

According to Eq.(\ref{B10}), the configuration average of the electron
density $\langle n(\rho ,s)\rangle $ is given by:
\begin{equation}
\left\langle n\left( \rho ,s\right) \right\rangle =\int d\rho _1p_0\left(
\rho _1\right) P\left( \rho _1,\rho \right) \,,  \label{CA1}
\end{equation}
where
\begin{equation}
P\left( \rho ^{\prime },\rho \right) =\eta \left( \rho ^{\prime }\right)
\Phi \left( \rho ^{\prime },\rho \right)  \label{CA2}
\end{equation}
is the diffusion function. According to Eq.(\ref{B11}), the diffusion
function is given by:
\begin{equation}
P=\frac 1s\sum\limits_{n=0}^\infty \eta \left( \frac Vs\right) ^n\,.
\label{CA3}
\end{equation}
Diagrammatically this series can be represented as depicted in Fig.\ref{FIG1}%
. In this picture every full dot represents a potential $V$, which depends
on disorder via the structur factor $\eta $. The expansion starts with a
single structural factor $\eta $, depicted by the empty dot. Note that all
disorder is comprised into the structure factor $\eta $.

In order to calculate the configuration average, we average the series (\ref
{CA3}) term by term. The resulting expansion can be depicted in the usual
form, as in impurity scattering problem (see Fig.\ref{FIG2}). One can see
from the picture, that the set of all diagrams can be decomposed into two
subsets, the set $S$ containing all diagrams connected with the empty point
and the set $F$ containing all other diagrams. Owing to this decomposition
configuration averaged diffusion function can be written as:
\begin{equation}
\left\langle P(\rho ^{\prime },\rho )\right\rangle =\int d\rho _1S\left(
\rho ^{\prime },\rho _1\right) F\left( \rho _1,\rho \right) \,,  \label{CA4}
\end{equation}
from which Eq.(\ref{C4}) follows. Note that $F$ is nothing but $\langle \Phi
\rangle $.

Again the class of diagrams contributing to $F$ can be decomposed into
reducible and irreducible diagrams. All irreducible diagrams can be
comprised into a function $\Pi $. Doing so, we obtain Eq.(\ref{C5}). Note,
that, although the introduction of the irreducible part $\Pi $ is parallel
to the introduction of the self-energy it has to be stressed, that its
physical interpretation is completely different. In the present context $\Pi
(\rho ^{\prime },\rho )$ is the true transition probability from $\rho
^{\prime }$ to $\rho $. It is $\Pi (\rho ^{\prime },\rho )$ that has to be
calculated and compared with the experimental situation and not the bar
transition probability, as suggested in many papers (see e.g. \cite{I5}%
). Diagrams, contributing to $\Pi $ are depicted in Fig.\ref{FIG3}. The set
of diagrams depicted in Fig. \ref{FIG3} can be generated by means of the
propagator $\Pi _{\tilde{\rho}}$, defined by the equation (\ref{C7}), using
Eq.(\ref{C6}).

The propagator $\Pi _{\tilde{\rho}}$ can also be used to generate the
irreducible block $S$. Diagrams contributing to $S$ are depicted in Fig.\ref
{FIG4}. From the picture it follows, that $S$ is given by Eq.(\ref{C8}).
\section{Transport coefficients}

\label{AP2}

According to Eq.(\ref{P7}), $v(\varepsilon ,s)$ is given by:
\begin{equation}
v\left( \varepsilon ,s\right) =\frac 12\omega ^2{\cal N}\left( \varepsilon
\right) \tilde{W}\left( q=0|\varepsilon \right) 
=\frac 12\omega ^2{\cal N}\left( \varepsilon \right) \nu \exp \left[ -\rho
_c\left( \varepsilon ,s\right) \right] \int d{\bf R}\frac 1{1+\exp \left[
2\alpha R-\rho _c\left( \varepsilon ,s\right) \right] }\,,  \label{T1}
\end{equation}
where:
\begin{equation}
f\left( \varepsilon ,s\right) \nu =\exp \rho _c\left( \varepsilon ,s\right)
\,.  \label{T2}
\end{equation}
We assume, that $\rho _c$ is large, so the integrals can be calculated in
the limit $\rho _c\to \infty $. This assumption is justified a posteriory by
explicit calculation of $\rho _c$. In limit $\rho _c\to \infty $, the
integrand reduces to a step function, so that the integrals can be
calculated easily. Doing so, we find:
\begin{equation}
v\left( \varepsilon ,s\right) =\frac 12\frac{S\left( d\right) }d\omega ^2%
{\cal N}\left( \varepsilon \right) \left[ \frac{\rho _c\left( \varepsilon
,s\right) }{2\alpha }\right] ^d\nu \exp \left[ -\rho _c\left( \varepsilon
,s\right) \right] \,,  \label{T3}
\end{equation}
where $S(d)$ is the solid angle and $d$ is the spatial dimension. From the
calculation and the result it is to be seen, that $\rho _c(\varepsilon ,s)$
is to be identified with the dimensionless critical hopping length at energy
$\varepsilon $.

The same procedure is applied to the calculation of $D(\varepsilon ,s)$,
defined by Eq.(\ref{P6}). Using the same approximations, we obtain:
\begin{equation}
D\left( \varepsilon ,s\right) =\frac 12\frac{S\left( d\right) }{d\left(
d+2\right) }\omega {\cal N}\left( \varepsilon \right) \left[ \frac{\rho
_c\left( \varepsilon ,s\right) }{2\alpha }\right] ^{d+2}\nu \exp \left[
-\rho _c(\varepsilon ,s)\right] \,.  \label{T4}
\end{equation}
Thus $D(\varepsilon ,s)$ and $v(\varepsilon ,s)$ differ only in
preexponential factors.

We anticipate that for $s=0$ we obtain:
\begin{equation}
\rho _c\left( \varepsilon ,0\right) =\frac{2\alpha }{\left[ \omega {\cal N}%
\left( \varepsilon \right) \right] ^{1/d}}\left[ \frac{2da}{S\left( d\right)
}\right] ^{1/d}\,,  \label{T5}
\end{equation}
where $a$ is a number and $S(d)$ is the solid angle. 
A 
different form for the diffusion coefficient can be found in Ref.\cite{5}.
The expression given in this paper differs from our expression essentially
only in that $\omega $ is replaced by the tailing parameter. In
looking on this difference one should, however, take into account, that the
model used in that paper is different. Furthermore, in Ref.\cite{5} the
transport coefficient are independent of $s$. It is therefore not completely
clear to which range of time they apply. Clearly, they can not determine the
evolution for all times.
\section{Derivation of self-consistency equation}

\label{AP3}

Here we derive the self-consistency equation (\ref{SC5}). In order to work
out explicitely the first order contribution to the self-consistency
equation (\ref{SC4}) we first need the first order correction to $\Pi $. It
is given by:
\begin{equation}
\Pi ^{(1)}\left( \rho ^{\prime },\rho ;s\right) =\int d\rho _1d\rho _2d\rho
_3{\cal N}\left( \varepsilon _3\right) {\tilde{w}}_{\rho _3}\left( \rho
^{\prime },\rho _1;s\right) \tilde{F}\left( \rho _1,\rho _2\right) {\tilde{w}%
}_{\rho _3}\left( \rho _2,\rho ;s\right) \,.  \label{AS1}
\end{equation}
If we insert the expression for ${\tilde{w}}_{\rho _3}$, we obtain:
\begin{eqnarray}
&&\Pi ^{(1)}\left( \rho \prime ,\rho ;s\right)=\int d\rho _1
\nonumber \\ &&\left\{ {\cal N}%
\left( \varepsilon \right) \tilde{W}\left( \rho \prime ,\rho ;s\right)
\left[ \tilde{F}\left( \rho ,\rho _1\right) -\tilde{F}\left( \rho \prime
,\rho _1\right) \right] \tilde{W}\left( \rho _1,\rho ;s\right)
-{\cal N}\left( \varepsilon _1\right) \tilde{W}%
\left( \rho ^{\prime },\rho _1;s\right) \left[ \tilde{F}\left( \rho _1,\rho
\right) -\tilde{F}\left( \rho ^{\prime },\rho \right) \right] \tilde{W}%
\left( \rho ,\rho _1;s\right) \right\} \,.  \label{AS2}
\end{eqnarray}
Thus, at this stage the self-consistency equation takes the form:
\begin{eqnarray}
0 &=&\int d{\bf q}d\varepsilon \prime d\varepsilon _1\left( \varepsilon
\prime -\varepsilon \right) \left\{ {\cal N}\left( \varepsilon \right)
\tilde{W}\left( 0|\varepsilon \prime ,\varepsilon ;s\right) \tilde{F}\left(
q|\varepsilon ,\varepsilon _1\right) \tilde{W}\left( q|\varepsilon
_1,\varepsilon ;s\right) 
-{\cal N}\left( \varepsilon \right) \tilde{W}\left(
q|\varepsilon ^{\prime },\varepsilon ;s\right) \tilde{F}\left( q|\varepsilon
^{\prime },\varepsilon _1\right) \tilde{W}\left( q|\varepsilon
_1,\varepsilon ;s\right) \right. \nonumber \\
&&\hspace{0.01in}\left.
-{\cal N}\left( \varepsilon _1\right) \tilde{W}\left(
0|\varepsilon ^{\prime },\varepsilon _1;s\right) \tilde{F}\left(
q|\varepsilon _1,\varepsilon \right) \tilde{W}\left( q|\varepsilon
,\varepsilon _1;s\right) 
+{\cal N}\left( \varepsilon _1\right) \tilde{W}%
\left( q|\varepsilon ^{\prime },\varepsilon _1;s\right) \tilde{F}\left(
q|\varepsilon ^{\prime },\varepsilon \right) \tilde{W}\left( q|\varepsilon
,\varepsilon _1;s\right) \right\} \,.  \label{AS3}
\end{eqnarray}
In order to eliminate $\tilde{F}$, we have to replace $\tilde{F}%
(q|\varepsilon ^{\prime },\varepsilon )=F(q|\varepsilon ^{\prime
},\varepsilon )-f(\varepsilon ,s)\delta (\varepsilon ^{\prime }-\varepsilon
) $, where $F$ is the effective medium approximation of the diffusion
propagator. Let us first work out the local contribution. Replacing $\tilde{F%
}$ by $f*\delta $, the first and the third term of Eq.(\ref{AS3}) cancel
each other. The 4th term is zero, taken into account that the delta function
of the effective medium is multiplied by its argument. Thus, when $\tilde{F}$
is replaced by $f(\varepsilon ,s)\delta (\varepsilon ^{\prime }-\varepsilon
) $, only the 2nd term of Eq.(\ref{AS3}) survives. To simplify this term
we take into account that
\begin{equation}\label{AS3a}
{\tilde W}(q|\epsilon;s)={\tilde W}(0|\epsilon,s)
\phi(\frac{q\rho_c(\epsilon,s)}{2\alpha}),
\end{equation}
where $\phi$ is a dimensionless function.
Consequently, we obtain 
\begin{eqnarray}
&-&\int d{\bf q}d\varepsilon ^{\prime }\left( \varepsilon ^{\prime
}-\varepsilon \right) {\cal N}\left( \varepsilon \right) f\left( \varepsilon
^{\prime },s\right) \left[ \tilde{W}\left( q|\varepsilon ^{\prime
},\varepsilon ;s\right) \right] ^2  
\approx -\int d{\bf q}\left[ \tilde{W}\left( {\bf q}|\varepsilon ,s\right)
\right] ^2\frac{\omega ^2}2f\left( \varepsilon ,s\right) {\cal N}\left(
\varepsilon \right) \nonumber \\
&=&-\frac{1}{2}\omega^2\left(\frac{2\alpha}{\rho_c(\epsilon,s)}\right)^d
{\cal N}(\epsilon)f(\epsilon,s){\tilde W}^2(0|\epsilon,s)\int d^dx\phi^2(x)
\,.  \label{AS4}
\end{eqnarray}
Here terms proportional to $\omega f^{-1}(\varepsilon ,s)df(\varepsilon
,s)/d\varepsilon \ll 1$ have been neglected.

Let us now focus on the contribution of the regular part of the diffusion
propagator to the self-consistency equation. Owing to the step functions in $%
F$ an $\tilde{W}$, the first and third term in Eq.(\ref{AS3}) are zero, when
$\tilde{F}$ is replaced by $F$. Thus we are left with:
\begin{equation}
\int d{\bf q}d\varepsilon \prime d\varepsilon _1\left( \varepsilon \prime
-\varepsilon \right) \left\{ -{\cal N}\left( \varepsilon \right) \tilde{W}%
\left( q|\varepsilon \prime ,\varepsilon ;s\right) F\left( q|\varepsilon
\prime ,\varepsilon _1\right) \tilde{W}\left( q|\varepsilon _1,\varepsilon
;s\right)
+{\cal N}\left( \varepsilon _1\right) \tilde{W}%
\left( q|\varepsilon ^{\prime },\varepsilon _1;s\right) F\left(
q|\varepsilon ^{\prime },\varepsilon \right) \tilde{W}\left( q|\varepsilon
,\varepsilon _1;s\right) \right\} \,.  \label{AS5}
\end{equation}
The range of integration in (\ref{AS5}) is determined by the step functions
in $F$ and $\tilde{W}(q|\varepsilon ^{\prime },\varepsilon ;s)=\theta
(\varepsilon ^{\prime }-\varepsilon )\theta (\omega -\varepsilon ^{\prime
}+\varepsilon )\tilde{W}(q|\varepsilon ^{\prime };s)$. Owing to these step
functions, the energy integrations extend at most over intervals of length $%
\omega $, so that the quasi-elastic approximation can again be applied
to the effective-transition probabilities and the density of states.
The diffusion propagator enetering Eq.(\ref{AS5}), however, can not be dealt 
with in this way since for arbitrary $q$ the derivatives of the diffusion
propagator are not small as compared to the diffusion propagator itself.
Therefore, another procedure is needed. To simplify this expression further
we consider Eq.(\ref{P1}). If terms small with respect to $\omega N'/N$ and 
$\omega {\tilde W}'(q|\epsilon,s))/{\tilde W}(q|\epsilon,s)$ are neglected
the function
\begin{equation}\label{AS5c}
\Phi(x|y',y)=\omega^2F(x\frac{2\alpha}{\rho_c(\omega y,s)}|\omega y',\omega
y){\tilde W}(\omega y,s){\cal N}(\omega y),
\end{equation}
can be introduced, that satisfies the equation
\begin{eqnarray}\label{AS5b}
\frac{s}{ {\tilde W}(\omega y,s){\cal N}(\omega y)\omega}
\Phi(x|y',y)=\delta(y'-y)+\int\limits_0^1 dy_1
[\Phi(x|y',y_1+y)\phi(x)-\Phi(x|y',y)\phi(0)].
\end{eqnarray}
Then the expression (\ref{AS5}) can be cast into the form
\begin{equation}\label{AS5h}
\omega{\tilde W}(\omega y)(\frac{2\alpha}{\rho_c(\omega y,s)})^d
\int d^dx\phi^2(x)\big[
-\int\limits_0^{1} dy'\int\limits_0^{y'} dy_1y'
\Phi(x|y'+\frac{\epsilon}{\omega},y_1+\frac{\epsilon}{\omega})
+ \int\limits_{0}^1\int\limits_0^{1-y_1}dy' y'
\Phi(x|y'+\frac{\epsilon}{\omega},\frac{\epsilon}{\omega})\big]
\end{equation}
For $s=0$ Eq.(\ref{AS5b}) does not contain any physical 
parameter. It only leads to the determination of  the function
\[\Phi_0(x|y',y)=\theta(y'-y)\Phi_0(x|y'-y),\]
which satisfies the equation
\begin{eqnarray}\label{AS5e}
0=\delta(y'-y)+\int\limits_0^1 dy_1
[\Phi_0(x|y',y_1+y)\phi(x)-\Phi_0(x|y',y)\phi(0)].
\end{eqnarray}
Provided, we restrict our consideration to small frequencies in deriving
the self-consistency equation we only
need to take into account the linear contribution of the function $\Phi$
with respect to $s/({\tilde W}(\omega y){\cal N}(\omega y) \omega)$.
Then, using again the smallness of the variation  of the effective 
transition probabilities and the density of states with respect to changes
of energy over intervals of length $\omega$, $\Phi$ can be 
approximated as
\begin{equation}\label{AS5f}
\Phi(x|y',y)=\Phi_0(x|y'-y)-\frac{s}{{\tilde W}(\omega y){\cal N}(\omega y)
\omega}\int\limits_y^{y'} dy_1 \Phi_0(x|y-y_1)\Phi_0(x|y_1-y).
\end{equation}
Using this expression the self-consistency equation takes the form
(\ref{SC5}), where the coefficients $a$ and $b$ are given by
\begin{equation}\label{AS5i}
a=\frac{1}{2}\frac{\int\limits d^dx\phi^2(x)
\int\limits_0^1(1-y)^2\Phi_0(x|y)}
{\int d^dx\phi^2(x)},
\end{equation}
\begin{equation}\label{AS6i}
b=\frac{1}{2}
\frac{\int d^dx\phi^2(x)\int_0^1 dy(1-y)^2\int_0^y dy_1\Phi_0(x |y-y_1)
\Phi_0(x|y_1)}{\int d^dx\phi^2(x)}.
\end{equation}
%


\section{Calculation of the energy relaxation speed for constant DOS}

\label{AP5}

To calculate the integral (\ref{CD1}), we again use Eq.(\ref{SC11}) to
change the integration variable from $s$ to $y=v/v_0$. Furthermore, for
convinience, we introduce the parameter $\tau =\Omega t$. Doing so, we
obtain:
\begin{equation}
\frac{dE\left( t\right) }{dt}=-v_0F\left( \tau \right) \,,  \label{CDOS1}
\end{equation}
where $F(\tau )$ is given by:
\begin{equation}
F\left( \tau \right) =\int_C\frac{dy}{2\pi i}\frac{1+\ln y}{\ln y}\exp
\left( \tau y\ln y\right) \,.  \label{CDOS2}
\end{equation}
After an integration by parts we obtain:
\begin{equation}
\frac{dF}{d\tau }=-\frac 1\tau \int_C\frac{dy}{2\pi i}\exp \left( \tau y\ln
y\right) \equiv -\frac 1\tau S\left( \tau \right) \,.  \label{CDOS3}
\end{equation}
To calculate $F(\tau )$, Eq. (\ref{CDOS3}) should be supplemented with the
condition $F(+\infty )=1$.

To evaluate $S\left( \tau \right) $, it is convenient to use the integration
contour $%
\mathop{\rm Im}%
(y\ln y)=0$, or, introducing polar coordinates $y=r\exp \left( i\theta
\right) $:
\begin{equation}
r=\exp \left( -\theta \cot \theta \right) \,.  \label{CDOS4}
\end{equation}
Substituting Eq.(\ref{CDOS4}) into Eq.(\ref{CDOS3}), we have:
\begin{equation}
S\left( \tau \right) =\int_{-\pi }^\pi \frac{d\theta }{2\pi }V\left( \theta
\right) e^{-\tau V\left( \theta \right) }\,,\;\;V\left( \theta \right) =%
\frac \theta {\sin \theta }e^{-\theta \cot \theta }\,.  \label{CDOS5}
\end{equation}
The value of the integral may be estimated by the saddle-point method,
looking for the maxima of the expression $\ln V\left( \theta \right) -\tau
V\left( \theta \right) $. The stationary point equation is:
\begin{equation}
V^{\prime }\left( \theta \right) \left[ \tau -\frac 1{V\left( \theta \right)
}\right] =0\,.  \label{CDOS6}
\end{equation}

As $\tau \gg 1$, the stationary point is $\theta _s=0$, and the asymptote of
Eq.(\ref{CDOS5}) is:
\begin{equation}
S\left( \tau \right) \approx \left( 2\pi e\tau \right) ^{-1/2}\exp \left(
-\tau /e\right) \,.  \label{CDOS7}
\end{equation}
On the other hand, if $\tau \ll 1$, we have two stationary points $\pm
\theta _s$, $V\left( \pm \theta _s\right) =1/\tau $, $\theta _s=\pi -\delta $%
, and for $\delta \ll 1$ the following equation may be obtained:
\[
\frac \pi \delta \exp \left( \frac \pi \delta -1\right) =\frac 1\tau \,,
\]
the solution for which at small $\tau $ is: $\pi /\delta ={\cal W}(e/\tau
)\approx \ln \left( e/\tau \right) -\ln \ln (e/\tau )\,.$The asymptotics of $%
S\left( \tau \right) $ turns out to be:
\begin{equation}
S\left( \tau \right) \approx \frac{\sqrt{2\pi }}{e\tau }\left[ \ln \left(
e/\tau \right) \right] ^{-2}\,.  \label{CDOS8}
\end{equation}
\section{The saddle-point approximation}

\label{SPA}

For energy dependent density of states it turns out to be difficult to
obtain explicite expressions for the time dependence of the energy
distribution function. Here, according to Eq.(\ref{PI0}), the
time-dependence has to be calculated from the equation:
\begin{equation}
P\left( \varepsilon ,\varepsilon _0,t\right) =\int\limits_{-i\infty
}^{+i\infty }\frac{ds}{2\pi iv\left( \varepsilon ,s\right) }\exp \left[
st-s\int\limits_\varepsilon ^{\varepsilon _0}\frac{d\varepsilon ^{\prime }}{%
v\left( \varepsilon ^{\prime },s\right) }\right] \,,  \label{SP0}
\end{equation}
where $v(\varepsilon ,s)$ is given by Eq. (\ref{SC11}) or (\ref{SC13}). Let
us try to integrate over $s$, using the saddle-point approximation. The
saddle-point position $s_0(\varepsilon ,\varepsilon _0,t)$ may be obtained
from the equation:
\begin{equation}
t=\int\limits_\varepsilon ^{\varepsilon _0}\frac{d\varepsilon ^{\prime }}{%
v\left( \varepsilon ^{\prime }\right) g\left( \varepsilon ^{\prime
},s_0\right) }-s_0\int\limits_\varepsilon ^{\varepsilon _0}\frac{%
d\varepsilon ^{\prime }}{v\left( \varepsilon ^{\prime }\right) g^2\left(
\varepsilon ^{\prime },s_0\right) }\frac{\partial g\left( \varepsilon
,s_0\right) }{\partial s_0}\,,  \label{SP1}
\end{equation}
where $g=v(\varepsilon ,s)/v(\varepsilon ,0)=\exp {\cal W}(s/\Omega
(\varepsilon ))$ was introduced. The expression for the diffusion propagator
in the saddle-point approximation may be written as:
\begin{equation}
P_L\left( \varepsilon ,\varepsilon _0,t\right) =\frac 1{\sqrt{4\pi D\left(
\varepsilon ,\varepsilon _0,s_0\right) }}\frac 1{v_0\left( \varepsilon
\right) g\left( \varepsilon ,s_0\right) }\exp \left[
-s_0^2\int\limits_\varepsilon ^{\varepsilon _0}\frac{d\varepsilon ^{\prime }%
}{v_0\left( \varepsilon ^{\prime }\right) g^2\left( \varepsilon ^{\prime
},s_0\right) }\frac{\partial g\left( \varepsilon ,s_0\right) }{\partial s_0}%
\right] \,,\,  \label{SP2}
\end{equation}
where:
\begin{equation}
D\left( \varepsilon ,\varepsilon _0,s_0\right) =-\frac{\partial ^2}{\partial
s_0^2}\int\limits_\varepsilon ^{\varepsilon _0}\frac{d\varepsilon ^{\prime
}\Omega \left( \varepsilon ^{\prime }\right) \ln g\left( \varepsilon
^{\prime },s_0\right) }{2v_0\left( \varepsilon ^{\prime }\right) }>0\,.
\label{SP3}
\end{equation}
Assuming $s_0/\Omega \ll 1$, we have: $g\left( \varepsilon ,s_0\right)
\simeq 1+s_0/\Omega -s_0^2/2\Omega ^2$,
\begin{equation}
s_0=\frac{t-T\left( \varepsilon ,\varepsilon _0\right) }{2D\left(
\varepsilon ,\varepsilon _0\right) }\,,\;\;D\left( \varepsilon ,\varepsilon
_0\right) =\int\limits_\varepsilon ^{\varepsilon _0}\frac{d\varepsilon
^{\prime }}{v_0\left( \varepsilon ^{\prime }\right) \Omega \left(
\varepsilon ^{\prime }\right) }\,,\;\;T\left( \varepsilon ,\varepsilon
_0\right) =\int\limits_\varepsilon ^{\varepsilon _0}\frac{d\varepsilon
^{\prime }}{v_0\left( \varepsilon ^{\prime }\right) }  \label{SP4}
\end{equation}
and, finally, the diffusion propagator:
\begin{equation}
P_L\left( \varepsilon ,\varepsilon _0,t\right) \simeq \frac 1{\sqrt{4\pi
D\left( \varepsilon ,\varepsilon _0\right) }}\frac 1{v\left( \varepsilon
\right) }\exp \left[ -\frac{\left( t-T\left( \varepsilon ,\varepsilon
_0\right) \right) ^2}{4D\left( \varepsilon ,\varepsilon _0\right) }\right]
\,.  \label{SP4A}
\end{equation}

At a given time $t$ the exponent in the above distribution function is
maximal at $\varepsilon =\varepsilon _m$, where $\varepsilon _m$ is given by
the condition $t=T(\varepsilon _m,\varepsilon _0)$, or:
\begin{equation}
t=\int\limits_{\varepsilon _m}^{\varepsilon _0}\frac{d\varepsilon ^{\prime }%
}{v_0\left( \varepsilon ^{\prime }\right) }\,.  \label{SP4B}
\end{equation}
Expanding the expression (\ref{SP4A}) around $\varepsilon =\varepsilon _m$,
we have gaussian distribution:
\begin{equation}
P_L\left( \varepsilon ,\varepsilon _0,t\right) \simeq \frac 1{\sigma \left(
\varepsilon _m,\varepsilon _0\right) \sqrt{2\pi }}\exp \left[ -\frac{\left(
\varepsilon -\varepsilon _m\left( \varepsilon _0,t\right) \right) ^2}{%
2\sigma ^2\left( \varepsilon _m,\varepsilon _0\right) }\right] \,,
\label{SP4C}
\end{equation}
with the dispersion:
\begin{equation}
\sigma ^2\left( \varepsilon _m,\varepsilon _0\right) =2v_0^2\left(
\varepsilon _m\right) D\left( \varepsilon _m,\varepsilon _0\right)
=2v^2\left( \varepsilon _m\right) \int\limits_{\varepsilon _m}^{\varepsilon
_0}\frac{d\varepsilon ^{\prime }}{v_0\left( \varepsilon ^{\prime }\right)
\Omega \left( \varepsilon ^{\prime }\right) }\,.  \label{SP4D}
\end{equation}
Note, that according to Eq.(\ref{SP4B}), the dispersion is time dependent.
If the density of states ${\cal N}(\varepsilon )$, and, consequently, $%
\Omega (\varepsilon )$ and $v_0(\varepsilon )$, are strongly varying
functions of $\varepsilon $, the integrals can be further simplified. In
this case we have:
\begin{equation}
\int\limits_{\varepsilon _m}^{\varepsilon _0}\frac{d\varepsilon ^{\prime }}{%
v\left( \varepsilon ^{\prime }\right) }\simeq \left[ \frac{dv\left(
\varepsilon _0\right) }{d\varepsilon _0}\right] ^{-1}-\left[ \frac{dv\left(
\varepsilon _m\right) }{d\varepsilon _m}\right] ^{-1}\,,  \label{SP5}
\end{equation}
\begin{equation}
\;\;\int\limits_{\varepsilon _m}^{\varepsilon _0}\frac{d\varepsilon ^{\prime
}}{v\left( \varepsilon ^{\prime }\right) \Omega \left( \varepsilon ^{\prime
}\right) }\simeq \left[ \frac{d\left[ v\left( \varepsilon _0\right) \Omega
\left( \varepsilon _0\right) \right] }{d\varepsilon _0}\right] ^{-1}-\left[
\frac{d\left[ v\left( \varepsilon _m\right) \Omega \left( \varepsilon
_m\right) \right] }{d\varepsilon _m}\right] ^{-1}\,,  \label{SP6}
\end{equation}
\begin{equation}
\int_{\varepsilon _m}^{\varepsilon _0}\frac{d\varepsilon ^{\prime }}{v\left(
\varepsilon ^{\prime }\right) \Omega \left( \varepsilon ^{\prime }\right) }%
\simeq \left[ \frac{dv\left( \varepsilon _0\right) \Omega \left( \varepsilon
_0\right) }{d\varepsilon _0}\right] ^{-1}-\left[ \frac{dv\left( \varepsilon
_m\right) \Omega \left( \varepsilon _m\right) }{d\varepsilon _m}\right]
^{-1}\,.  \label{SP7}
\end{equation}

The applicability condition for the saddle-point method is:
\begin{equation}
D^3\left( \varepsilon ,\varepsilon _0,s_0\right) \gg \left[ \frac{\partial ^3%
}{\partial s_0^3}\int\limits_\varepsilon ^{\varepsilon _0}d\varepsilon
^{\prime }\,\frac{\Omega \left( \varepsilon ^{\prime }\right) \ln g\left(
\varepsilon ^{\prime },s_0\right) }{2v\left( \varepsilon ^{\prime }\right) }%
\right] ^2\,.  \label{SP8}
\end{equation}
Note, that this condition restricts the applicablity of the
saddle-point approximation to times, that are not too large. Also, from 
Eqs.(\ref{SP4}) one can see, that this approximation is
invalid when $\varepsilon $ is close enough to the initial point $%
\varepsilon _0$, --- here $\left| s_0\right| $ becomes large, and it is not
possible to expand over $s_0$. So, the initial stage of the evolution, when
the entire distribution is concentrated near $\varepsilon _0$, has to be
investigated separately too.

\section{Calculation of the momenta for large times}

\label{AP6}

Here we present details on the calculation of the momenta of the
distribution function for large times and for the exponential density of
states. In our calculation we take into account that the density of states
decays with decreasing energy.

Let us change the integration variable in (\ref{DLT4}), $\rho \rightarrow z=%
{\cal W}\left( s\rho e^\rho \right) $, $\rho ={\cal W}\left(
s^{-1}ze^z\right) $. As $\rho \gg 1$, we have, introducing $\rho _s={\cal W}%
\left( s^{-1}\right) $, $s=\rho _s^{-1}\exp \left( -\rho _s\right) $:
\[
\rho ={\cal W}\left( \rho _sze^{\rho s+z}\right) \approx \rho _s+z+\ln z\,,
\]
as $\left| z\right| \ll \left| \rho _s\right| $ and $\left| \ln z\right| \ll
\left| \rho _s\right| $. Both inequalities holds within actual region of
integration. We can write now:
\begin{equation}
\chi _{nL}(s)=-\frac 1s\int\limits_0^\infty dz\,\left[ -\ln \rho _s-\frac{%
z+\ln z}{\rho _s}\right] ^n\frac \partial {\partial z}\exp \left[ -\frac 1{%
\bar\omega \rho _s^2}z\left( 1+\frac 12z\right) \right] \,.  \label{MM4}
\end{equation}
Something more about the approximation we have chosen: every extra power of $%
z/\rho _s$ means an extra power of $\sqrt{\bar\omega }$, which can be 
neglected.
An extra power of $z$ means the multiplier $\sim \sqrt{\bar\omega }\rho _s$. 
All
our considerations are valid if $\bar\omega \rho _s\ll 1$ only. Therefore, 
we
have the reason to omit extra powers of $z/\rho _s\sim \sqrt{\bar\omega }$ 
in
the following, but to keep some extra powers of 
$z^{-1}\sim 1/\sqrt{\bar\omega }\rho _s\gg \sqrt{\bar\omega }$.

Thus we have, for example:
\begin{eqnarray}
&&\chi _1(s)=-\frac 1s\ln \rho _s+\frac 1{s\rho _s}\int\limits_0^\infty
dz\,\left( z+\ln z\right) \frac d{dz}\exp \left[ 
-\frac 1{\bar\omega \rho _s^2}%
z\left( 1+\frac 12z\right) \right]  \nonumber \\
&=&-\frac 1s\ln \rho _s+\frac{\sqrt{\bar\omega }}s\int\limits_0^\infty
dy\,\left( y+\frac{\ln \left( \sqrt{\bar\omega }\rho _s\right) }
{\sqrt{\bar\omega }%
\rho _s}+\frac{\ln y}{\sqrt{\bar\omega }\rho _s}\right) 
\frac d{dy}\exp \left( -%
\frac{y^2}2-\frac y{\sqrt{\bar\omega }\rho _s}\right) ,
\end{eqnarray}\label{MM5}
where
\[y=\frac z{\sqrt{\bar\omega }\rho _s}\,.\]

Within three leading orders on the parameter $1/\sqrt{\bar\omega }\rho _s$ 
we have:
\begin{equation}
\chi _{1L}(s)\approx -\frac{\ln \rho _s}s-\sqrt{\frac{\pi \bar\omega }2}
\frac 1s-
\frac{\ln \left( \sqrt{\bar\omega }\rho _s\right) }
{s\rho _s}+\left( 1-\frac{\ln
2-\gamma }2\right) \frac 1{s\rho _s}+\sqrt{\frac \pi {8\bar\omega }}
\frac 1{%
s\rho _s^2}\,\,,  \label{MM6}
\end{equation}
which can be written, using $\rho _s+\ln \rho _s=\ln \left( 1/s\right) $
also as:
\begin{equation}
\chi _{1L}(s)=-\frac 1s\ln \ln \frac{b\sqrt{\bar\omega }}s
-\sqrt{\frac{\pi
\bar\omega }2}\frac 1s
+\sqrt{\frac \pi {8\bar\omega }}
\frac 1{s\rho _s^2}\,,\,b=
\sqrt{2}e^{-1-\gamma /2}\,,  \label{MM7}
\end{equation}
where $\gamma $ is the Euler's constant. 
In the time representation we have:
\[
\chi _1(t)=\int_C\frac{ds}{2\pi i}\chi _{1L}(s)e^{st}=
-\int_C\frac{ds}{2\pi
is}\ln \ln \frac{b\sqrt{\bar\omega }}
se^{st}-\sqrt{\frac{\pi \bar\omega }2}+\sqrt{
\frac \pi {8\bar\omega }}\int_C\frac{ds}
{2\pi is\ln ^2\frac{\sqrt{\bar\omega }}s}%
e^{st}\,.
\]
At large $t$, $b\sqrt{\bar\omega }t\gg 1$, the first integral, after the
substitution: $st=y$, may be expanded as:
\begin{eqnarray*}
&&\int_C\frac{ds}{2\pi is}\ln \ln \frac{b\sqrt{\bar\omega }}
se^{st}=\int_C\frac{%
dy}{2\pi iy}\ln \left[ \ln b\sqrt{\bar\omega }t
-\ln y\right] e^y=\ln \ln b\sqrt{%
\bar\omega }t \\
&&-\frac 1{\ln b\sqrt{\bar\omega }t}
\int_C\frac{dy}{2\pi iy}e^y\ln y-\frac 1{%
2\ln ^2b\sqrt{\bar\omega }t}\int_C
\frac{dy}{2\pi iy}e^y\ln ^2y-\dots \\
&=&\ln \ln b\sqrt{\bar\omega }t+\frac \gamma 
{\ln b\sqrt{\bar\omega }t}+\frac{\pi
^2/6-\gamma ^2}{2\ln ^2b\sqrt{\bar\omega }t}\,.
\end{eqnarray*}
As for the second integral, one may restrict oneself with the leading term
only. As a result, we have:
\begin{equation}
\chi _1(t)=-\ln \ln \tilde{b}\sqrt{\bar\omega }t
-\sqrt{\frac{\pi\bar\omega }2}+%
\sqrt{\frac \pi {8\bar\omega }}\frac 1{\ln ^2\tilde{b}
\sqrt{\bar\omega }t}+\dots
\,,\;\tilde{b}=\sqrt{2}e^{-1+\gamma /2}\,.  \label{MM8}
\end{equation}

Let us consider now the second momentum:
\begin{equation}
\chi _{2L}(s)=\int\limits_0^\infty \frac{d\rho }\rho \,\ln ^2\rho \,\phi
_L\left( \rho ;s\right) \equiv \left\langle \ln ^2\rho \right\rangle \,.
\label{MM9}
\end{equation}
Again, we change the integration variable to $z={\cal W}\left( s\rho e^\rho
\right) $, write $\rho \gg 1$ as $\rho =\rho _s+z+\ln z$, and set the lower
limit of the integration to be zero. Restricting ourselves with the lowest
order in $z/\rho _s\sim \sqrt{\bar\omega }$, we set $\ln \rho =\ln \rho
_s+(z+\ln z)/\rho _s$, and:
\[
\frac{d\rho }\rho \phi _L=-\frac{dz}s\,\frac \partial {\partial z}\exp
\left[ -\frac 1{\bar\omega \rho _s^2}z\left( 1+\frac 12z\right) \right] \,.
\]
Taking into account:
\[
\chi _{1L}(s)=-\frac 1s\ln \rho _s-\left\langle \frac{z+\ln z}{\rho _s}%
\right\rangle \,,
\]
we can write:
\begin{equation}
\chi _{2L}(s)=s\chi _{1L}^2(s)+\left\langle \left( \frac{z+\ln z}{\rho _s}%
\right) ^2\right\rangle -\left\langle \frac{z+\ln z}{\rho _s}\right\rangle
^2\,.  \label{MM10}
\end{equation}
Then we have:
\begin{eqnarray*}
\left\langle \left( \frac{z+\ln z}{\rho _s}\right) ^2\right\rangle &=&-\frac %
1{s\rho _s^2}\int\limits_0^\infty dz\,(z+\ln z)^2\frac d{dz}\exp \left[ -%
\frac 1{\bar\omega \rho _s^2}z\left( 1+\frac 12z\right) \right] \\
&=&-\frac {\bar\omega} s\int\limits_0^\infty dy\,\left( y+
\frac{\ln \left( \sqrt{\bar\omega }\rho _s\right) }
{\sqrt{\bar\omega }\rho _s}+\frac{\ln y}{\sqrt{\bar\omega }%
\rho _s}\right) ^2\exp \left( -\frac{y^2}2-\frac y{\sqrt{\bar\omega }\rho _s}%
\right) \,.
\end{eqnarray*}
Expanding this expressions in power series on $1/\sqrt{\bar\omega }\rho _s$, 
and
keeping three leading powers, it may be easily obtained:
\begin{eqnarray*}
&&\left\langle \left( \frac{z+\ln z}{\rho _s}\right) ^2\right\rangle =\frac{%
2\bar\omega }s+\frac{\sqrt{2\pi \bar\omega }\ln \left( \sqrt{\bar\omega }
\rho _s\right)
}{s\rho _s}-\frac{\ln 2+\gamma }2\frac{\sqrt{2\pi \bar\omega }}
{s\rho _s}+\frac{%
\ln ^2\left( \sqrt{\bar\omega }\rho _s\right) }{s\rho _s^2} \\
&&-2\left( 1-\frac{\ln 2-\gamma }2\right) \frac{\ln \left( 
\sqrt{\bar\omega }
\rho _s\right) }{s\rho _s^2}+\left[ \frac{\pi ^2}{24}-\ln 2+\gamma +\left(
\frac{\ln 2-\gamma }2\right) ^2\right] \frac 1{s\rho _s^2}\,.
\end{eqnarray*}
>From Eq. (\ref{MM6}) we have:
\[
\left\langle \frac{z+\ln z}{\rho _s}\right\rangle =
\sqrt{\frac{\pi \bar\omega }2}
\frac 1s+\frac{\ln \left( 
\sqrt{\bar\omega }\rho _s\right) }{s\rho _s}-\left( 1-%
\frac{\ln 2-\gamma }2\right) \frac 1{s\rho _s}-
\sqrt{\frac \pi {8\bar\omega }}%
\frac 1{s\rho _s^2}\,.
\]
So, from Eq. (\ref{MM10}) we obtain:
\begin{equation}
\chi _{2L}(s)=\chi _{1L}^2(s)+\left( 2-\frac \pi 2\right) +\left( 1-\ln
2\right) \frac{\sqrt{2\pi \bar\omega }}{s\rho _s}
+\left( \frac{\pi ^2}{24}+\frac %
\pi 2-1\right) \frac 1{s\rho _s^2}\,.  \label{MM11}
\end{equation}

However, we want to calculate {\em central} momentum:
\begin{equation}
\mu _2(t)=\chi _2(t)-\chi _1^2(t)\,,  \label{MM12}
\end{equation}
or, in the Laplace representation:
\begin{eqnarray}
\mu _{2L}(s) &=&\chi _{2L}(s)-\left( \chi _1^2\right) _L(s)\,,  \nonumber \\
\left( \chi _1^2\right) _L(s) &\equiv &\int\limits_0^\infty dt\,e^{st}\chi
_1^2(t)=\int\limits_{-i\infty +0}^{+i\infty +0}\frac{ds_1}{2\pi i}\chi
_{1L}(s_1)\chi _{1L}(s-s_1)\,.  \label{MM13}
\end{eqnarray}
Taking the expression for $\chi _{1L}$ from Eq. (\ref{MM7}), we have:
\begin{eqnarray}
&&\left( \chi _1^2\right) _L(s)=\int\limits_{-i\infty +0}^{+i\infty +0}\frac{%
ds_1}{2\pi i}\frac{\ln \ln \frac{b\sqrt{\bar\omega }}s\ln 
\ln \frac{b\sqrt{%
\bar\omega }}{s-s_1}}{s_1\left( s-s_1\right) }+
\frac{\sqrt{2\pi\bar\omega }}s\ln
\ln \frac{b\sqrt{\bar\omega }}s+\frac{\pi \bar\omega }{2s}  \nonumber \\
&&-\int\limits_{-i\infty +0}^{+i\infty +0}\frac{ds_1}{2\pi i}\frac{\ln \ln
\frac{b\sqrt{\bar\omega }}{s_1}}{s_1\left( s-s_1\right) 
\ln ^2\frac{b\sqrt{%
\bar\omega }}{s-s_1}}-\frac \pi {2s\ln ^2\frac{b\sqrt{\bar\omega }}s}
+\dots \,.
\label{MM14}
\end{eqnarray}
Let us consider the first integral. After changing the integration variable,
$s_1=sx$, we have:
\begin{eqnarray*}
&&\frac 1s\ln ^2\ln \frac{b\sqrt{\bar\omega }}s+
\frac 2s\ln \frac{b\sqrt{\bar\omega }%
}s\int_C\frac{dx}{2\pi i}\frac{\ln \left( 1-\ln (1-x)/\ln \left( \frac{b%
\sqrt{\bar\omega }}s\right) \right) }{x(1-x)} \\
&&+\int_c\frac{dx}{2\pi i}\frac{\ln \left( 1-\ln x/\ln \left( \frac{b\sqrt{%
\bar\omega }}s\right) \right) \ln 
\left( 1-\ln (1-x)/\ln \left( \frac{b\sqrt{%
\bar\omega }}s\right) \right) }{x(1-x)}\,.
\end{eqnarray*}
Closing the integration contour to the left, one can easily prove, that the
first intehral in the above expression is zero. The second one at small $s$
may be expanded into powers of $1/\ln s$. In the leading order we have:
\[
\frac 1s\ln ^2\ln \frac{b\sqrt{\bar\omega }}s
+\frac 1{s\ln ^2\frac{b\sqrt{\bar\omega
}}s}\int_C\frac{dx}{2\pi i}\frac{\ln x\ln (1-x)}{x(1-x)}\,.
\]
Performing integration by part in the integral, and then deforming the
integration contour to make it run along the lower and upper shores of the
cut $\left( -\infty ,0\right) $ of the integration function:
\begin{eqnarray*}
\int_C\frac{dx}{2\pi i}\frac{\ln x\ln (1-x)}{x(1-x)} &=&-\frac 12\int_C\frac{%
dx}{2\pi i}\ln ^2x\frac d{dx}\frac{\ln (1-x)}{1-x}=\frac 1{2\pi }%
\int\limits_0^\infty dx\frac{\ln (1+x)-1}{(1+x)^2}%
\mathop{\rm Im}%
\ln ^2(x+i0) \\
&=&\int\limits_0^\infty dx\frac{\left[ \ln (1+x)-1\right] \ln x}{(1+x)^2}=%
\frac{\pi ^2}6\,.
\end{eqnarray*}
The evaluation of the second integral in Eq. (\ref{MM14}) gives:
\[
-2\frac{\ln \ln \frac{b\sqrt{\bar\omega }}s}
{s\ln ^2\frac{b\sqrt{\bar\omega }}s}%
+O\left( \frac 1{s\ln ^4\frac{b\sqrt{\bar\omega }}s}\right)
\]
As a result, one can write:
\begin{equation}
\left( \chi _1^2\right) _L(s)=s\chi _{1L}^2(s)+\frac{\pi ^2}{6s\rho _s^2}%
+\cdots  \label{MM15}
\end{equation}
Thus, one have the second central momentum in the Laplace representation to
be:
\begin{equation}
\mu _{2L}(s)=(2-\frac \pi 2)\frac {\bar\omega} s
+(1-\ln 2)\frac{\sqrt{2\pi \bar\omega }%
}{s\rho _s}+\left( \frac \pi 2-\frac{\pi ^2}8-1\right) \frac 1{s\rho _s^2}\,,
\label{MM16}
\end{equation}
and the time dependence of the second central momentum:
\begin{equation}
\mu _2(t)=\left( 2-\frac \pi 2\right) \bar\omega 
+\left( 1+\ln 2\right) \frac{%
\sqrt{2\pi \bar\omega }}{\ln t}\,.  \label{MM17}
\end{equation}
Expression for the third central momentum $\mu _3\left( t\right) $ (\ref
{DLT7}) may be obtained quite analogously.

\section{Asymptotic form of the distribution function at large times}

\label{AP7}

One can write the large times distribution function (\ref{DLT3}), after the
replacement of the integration variable $s\rightarrow z={\cal W}\left( s\rho
e^\rho \right) $as follows:
\begin{equation}
\phi (\varepsilon ,t)=\frac \partial {\partial t}\psi (\varepsilon
,t)
\end{equation}
\begin{equation}
\psi (\varepsilon ,t)=\rho e^\rho \int\limits_{-i\infty +\delta
}^{+i\infty +\delta }\frac{dz}{2\pi iz^2}(1+z)\frac \partial {\partial
\varepsilon }\exp \left[ -z+t\frac z\rho e^{z-\rho }
-\frac 1{\bar\omega \rho ^2}%
z\left( 1+\frac 12z\right) \right] .  \label{LT1}
\end{equation}
One can try to calculate the latter integral within the saddle point
approximation. The saddle point equation is:
\[
f^{\prime }(z)=-1+t\frac{z+1}\rho e^{z-\rho }-
\frac 1{\bar\omega \rho ^2}%
(z+1)=0\,.
\]
Assuming $\left| z\right| \ll \bar\omega \rho ^2$ 
(remind that $\omega \rho
^2\gg 1$), we can neglect the third term in the above equation, and to
obtain:
\[
z_c+1={\cal W}\left( \frac \rho te^{\rho +1}\right) \,.
\]
As we shall see later, the actual values of parameters $\rho $, $t$,
corresponding to the body of the distribution function, obeys the inequality
$\rho e^\rho /t\gg 1$, and, therefore, $z_c\gg 1$. But this leads to the
saddle point parameter, $\left| f^{\prime \prime \prime }(z_c)/\left[
f^{\prime \prime }(z_c)\right] ^{3/2}\right| =\left|
(z_c+3)/(z_c+2)^{3/2}\right| \approx 1$, to be of the order of 1, instead of
being much less. Let us note, however, that the function in the exponent in
Eq. (\ref{LT1}) may be represented as a sum of two terms, of which the first
one, $-z+tz/\rho \exp \left( z-\rho \right) $, ensures the function in the
integral to have a maximum at $z=z_c$ of the width $\sim 1$, while the
second one, $\left( \bar\omega \rho ^2\right) ^{-1}
z\left( 1+z/2\right) $, may
be supposed to be almost constant within this peak, --- it varies
essentially at scales $\Delta z\sim \sqrt{\bar\omega }\rho \gg 1$ 
within the
actual interval of system's parameters. Thus, one can replace that second
term in the exponent by its value at some $z_0=z_c+a$, $\left| a\right| \sim
1$. After this, the integration may be easily performed, returning back to
the initial integration variable $s$:
\[
\psi (\varepsilon ,t)\approx -\rho \frac \partial {\partial \rho }\exp
\left\{ -\frac 1{\bar\omega \rho ^2}
{\cal W}\left( s_0\rho e^\rho \right) \left[
1+\frac 12{\cal W}\left( s_0\rho e^\rho \right) \right] \right\}
\int\limits_{-i\infty +\delta }^{+i\infty +\delta }\frac{ds}{2\pi is^2}%
e^{st}\,,
\]
where the last integral is equal to simply $t$, and, taking into account $%
z_0\gg 1$, $s_0=z_0/\rho \exp \left( z_0-\rho \right) \approx c/t$, where $%
c\sim 1$ will be obtained later. Performing differentiation with respect to $%
t$, see Eq. (\ref{LT1}), and neglecting $z_0/\bar\omega \rho ^2\ll 1$, we 
obtain
finally:
\begin{equation}
\phi (\varepsilon ,t)\approx -\rho \frac \partial {\partial \rho }\exp
\left\{ -\frac 1{\bar\omega \rho ^2}{\cal W}
\left( \frac ct\rho e^\rho \right)
\left[ 1+\frac 12{\cal W}\left( \frac ct\rho e^\rho \right) \right] \right\}
\,.  \label{LT2}
\end{equation}

To obtain $c$, let us calculate $\chi _1(t)=\bar{\varepsilon}(t)$ with the
distribution function (\ref{LT2}). We have:
\[
\chi _1(t)=-\int \frac{d\rho }\rho \phi \left( \rho ,t\right) \ln \rho
\,=\int dz\,\ln \rho (z)\,\frac \partial {\partial z}\exp \left[ -\frac 1{%
\bar\omega \rho ^2(z)}z
\left( 1+\frac 12z\right) \right] \]
\[\rho (z)={\cal W}%
\left( \frac tcze^z\right) \,,
\]
where the replacement of the integration variable $\rho \rightarrow z={\cal W%
}\left( c\rho e^\rho /t\right) $ was done. Assuming $\rho \gg 1$, one can
write, using ${\cal W}\left( x\right) \approx x-\ln x$ as $\left| x\right|
\gg 1$, that $\rho (z)\approx {\cal W}\left( t/c\right) +z+\ln z\equiv \rho
_t+z+\ln z$ (the condition $z\sim \sqrt{\bar\omega }
\rho _t\ll \rho _t={\cal W}%
\left( t/c\right) $ was used). In the leading order on $\sqrt{\bar\omega }$, 
one
can replace $\rho (z)$ within the exponent with $\rho _t$, and $\ln \rho
(z)\rightarrow \ln \rho _t+(z+\ln z)/\rho _t$. The limits of integration by $%
z$ are to be set $\left( 0,+\infty \right) $. Then we have:
\begin{equation}
\chi _1(t)=-\ln \rho _t-\int\limits_0^\infty dz\,\frac{z+\ln z}{\rho _t}%
\frac \partial {\partial z}\exp \left[ -\frac 1{\bar\omega \rho _t^2}z
\left( 1+%
\frac 12z\right) \right] \,.  \label{LT3}
\end{equation}
This latter integral, analogously to one in Eq. (\ref{MM5}), may be
evaluated as an expansion on small parameter $1/\sqrt{\bar\omega }\rho $:
\begin{equation}
\chi _1\left( t\right) =-\ln \ln \left( \sqrt{2}e^{-1-\gamma /2}\frac{\sqrt{%
\bar\omega }t}c\right) -\sqrt{\frac{\pi \bar\omega }2}+\cdots  \label{LT4}
\end{equation}
Comparing Eqs.(\ref{LT4}) and (\ref{DLT5}), one can see, that $c=\exp \left(
-\gamma \right) $.\\

\begin{multicols}{2}
\narrowtext

\end{multicols}

\end{document}